\documentclass[preprint2]{aastex}
\usepackage{lscape}

\def\llog{{\boldmath{ \log (L_{\rm x}/{\rm erg\, s^{-1}})}}}

\begin{document}

\title{The Mass of the Black Hole in LMC X-3$^{\dagger}$}

\author{Jerome A. Orosz}
\affil{Department of Astronomy, San Diego State University,
5500 Campanile Drive, San Diego, CA 92182-1221}
\email{orosz@sciences.sdsu.edu}

\author{James F. Steiner, Jeffrey E. McClintock}
\affil{Harvard-Smithsonian Center for Astrophysics, 60 Garden Street,
Cambridge, MA 02138}
\email{jsteiner@cfa.harvard.edu,jem@cfa.harvard.edu}

\author{Michelle M. Buxton, Charles D. Bailyn}
\affil{Department of Astronomy, Yale University, PO Box 208101,
New Haven, CT 06520-8101}
\email{michelle.buxton@yale.edu,charles.bailyn@yale.edu}

\author{Danny Steeghs}
\affil{Department of Physics, University of Warwick, Coventry, CV4 7AL, UK
and
Harvard-Smithsonian Center for Astrophysics, 60 Garden Street, Cambridge, MA
02138}
\email{D.T.H.Steeghs@warwick.ac.uk}

\author{Alec Guberman}
\affil{Byram Hills High School, 12 Tripp Lane, Armonk, NY 10504
and 
Department of Physics and Astronomy,
Stony Brook University, Stony Brook, NY 11794-3800}
\email{alec.guberman@stonybrook.edu}

\and

\author{Manuel A. P. Torres}
\affil{SRON Netherlands Institute for Space Research, 
Sorbonnelaan 2, 3584 CA Utrecht, the Netherlands
and
Harvard-Smithsonian Center for Astrophysics, 60 Garden Street, Cambridge, MA
02138}
\email{M.Torres@sron.nl}

\altaffiltext{$\dagger$}{Based on observations made with
the Magellan 6.5m Clay
telescope at Las Campanas Observatory of the Carnegie Institution
and on 
data products from 
observations made with ESO Telescopes at the Paranal Observatory
under programme ID 074.D-0143}

\begin{abstract}
We analyze a large set of new and archival photometric and
spectroscopic observations of LMC X-3 to arrive at a
self-consistent dynamical model for the system.  Using
echelle spectra obtained with the MIKE instrument on the 6.5m 
Magellan Clay telescope and the UVES instrument on the
second 8.2m Very Large Telescope we find a velocity semiamplitude for the
secondary star of $K_2=241.1\pm 6.2$ km s$^{-1}$, where the uncertainty
includes an estimate of the systematic error caused by X-ray heating. 
Using the spectra, we also find
a 
projected rotational velocity of $V_{\rm rot}\sin i=118.5\pm
6.6$ km s$^{-1}$.  From an analysis of archival $B$ and $V$ light curves
as well as new $B$ and $V$ light curves from the SMARTS 1.3m telescope,
we find an inclination of $i=69.84\pm 0.37^{\circ}$ for
models that do not include X-ray heating and an inclination of
$i=69.24\pm 0.72^{\circ}$
for models that incorporate X-ray heating. Adopting the latter
inclination measurement, we find masses of $3.63\pm 0.57\,M_{\odot}$
and $6.98\pm 0.56\,M_{\odot}$ for the companion star and the black
hole, respectively.  
We briefly compare our results with earlier work and discuss some of their implications.
\end{abstract}

\section{Introduction}

LMC X-3 was the second black hole, after Cygnus X-1, to be established
via dynamical observations (Cowley et al.\ 1983).  The large mass
function of the host binary, $f(M) = 2.3 \pm 0.3~M_{\odot}$, the
absence of X-ray eclipses, and the estimated mass of the B3 V
secondary ($4-8~M_{\odot}$) allowed Cowley et al.\ to conclude that
``the most probable mass'' of the black hole is $M \sim 9 M_{\odot}$,
with a plausible lower limit of $M > 7~M_{\odot}$.  A subsequent study
of the 1.7-day optical light curve constrained the inclination $i$ to
lie in the range $\sim64^{\circ}$ to $70^{\circ}$, and it also
provided a firm lower limit on the mass of the primary, $M >
3.5~M_{\odot}$ (Kuiper et al.\ 1988), thereby confirming that it is
indeed a black hole (e.g., Kalogera \& Baym 1996).  Recently, in the
context of a far-UV study of LMC X-3 (Song et al.\ 2010), and using
much of the data discussed herein, we reported a precise ephemeris and
radial velocity amplitude for the secondary.  

It is a challenge to establish a reliable dynamical model of the LMC
X-3 system because both the optical counterpart and the X-ray source
are highly variable.  In addition to the $\approx 0.15$ mag
ellipsoidal variations (Kuiper et al.\ 1988), the optical flux varies
chaotically by up to $\approx 1$ mag (Brocksopp et al.\ 2001), while
the {\it RXTE} PCA shows that the X-ray luminosity ranges over a
factor of at least 2,500 (Smale \& Boyd 2012).

Furthermore, the optical variability (which is most relevant for a
dynamical study) is not simply slaved to the X-ray variability as it is
in most transient black hole binaries: In a typical transient system,
the optical emission is generated directly and promptly by X-ray
reprocessing in the outer accretion disk or in the X-ray-heated face of
the secondary (van Paradijs \& McClintock 1995).  While both of these
effects do occur in LMC X-3, the extra complicating factor for this
source is the variable mass accretion rate in the outer disk that
produces an additional, large-amplitude component of optical and
(delayed) X-ray variability (Brocksopp et al.\ 2001; Steiner et
al. 2013).

The absorbing column depth to LMC X-3 is stable and extraordinarily
small, $N_{\rm H} = 3.8_{-0.7}^{+0.8} \times 10^{20}$~cm$^{-2}$ (Page et
al.\ 2003) and, correspondingly, the optical extinction is low, $A_{\rm
V} \approx 0.2$~mag.  Also, the X-ray spectrum is normally
disk-dominated (Wilms et al.\ 2001; Steiner et al.\ 2010).  This
combination of low absorption and a soft spectrum, combined with the
persistence of the source, has made LMC X-3 a touchstone for testing
accretion disk models (Davis et al.\ 2006; Kubota et al.\ 2010; Straub
et al.\ 2011), and for stringently testing the stability of the inner
disk radius as a foundation for the measurement of black hole spin
(Steiner et al.\ 2010).

Davis et al.\ (2006) have used the continuum-fitting method to estimate
the dimensionless spin parameter of the black hole primary in LMC X-3 to
be $a_* \lesssim 0.3$, a value that depends strongly on the provisional
estimates of the black hole mass $M$ and inclination $i$ that are
discussed above (Cowley et al.\ 1983; Kuiper et al.\ 1988).  In this
paper, which is based on an extensive collection of spectroscopic and
photometric data, we present a new dynamical model for LMC X-3.  In a
companion paper (Steiner et al. 2014), we make use of the precise values
of $M$ and $i$ reported herein, the secure distance to LMC X-3, and
extensive archival X-ray data (Steiner et al.\ 2010) to obtain a firm
estimate of the black hole's spin.

This paper is organized as follows.  We discuss the new
and archival spectroscopic observations in \S2 and the new and
archival photometric observations in \S3.  In \S4 we discuss
the analysis of the spectra including the measurement of the
radial velocities, the rotational velocity, and the detection 
of emission lines in some of the spectra.
In \S5 we first present our 
extensive collection of light 
curve data, and we then present our dynamical 
model.  We discuss several topics and offer our conclusions in \S6.

\section{Observations}\label{specsec}

\subsection{Magellan Spectroscopic Observations}

Our echelle spectra obtained using the Magellan Inamori Kyocera Echelle
(MIKE) spectrograph (Bernstein et al.\ 2002) and the 6.5 m Magellan Clay
telescope at Las Campanas Observatory (LCO) were previously discussed in
Song et al.\ (2010), but for completeness many of the details are given
here.  
Fifty-four spectra of the optical counterpart of LMC X-3 (Warren \& Penfold 1975) 
along with the spectra 
of  several flux standards
and
spectral-comparison stars were obtained on the nights of 2005 January
20--24, 2007 December 20--21, and 2008 February 27 through 2008 March 1.
MIKE was used in the standard dual-beam mode with a $1\farcs
0\times5\farcs 0$ slit and the $2\times 2$ binning mode.
The $1^{\prime\prime}$ slit width was well-matched to the seeing, which 
was between 0\farcs 7 and 0\farcs 9 for the 2005 run, 0\farcs 8 to 1\farcs 2 for the 2007
run, and 0\farcs 6 and 1\farcs 2 for
the 2008 run.  
The wavelength calibration was provided by exposures of a
ThAr lamp, obtained before and after each
pair of observations of the object.
We used data from the blue arm, which has a wavelength
coverage of 3430--5140 \AA\ and resolving power of $R=33,000$.

The results presented in Song et al.\ (2010) were based on reductions of
the MIKE spectra carried out with an IDL-based pipeline written by Scott
Burles\footnote{http://web.mit.edu/$\sim$burles/www/MIKE/}.  Since that
time we have found a possible problem with the heliocentric wavelength
corrections.  Also, the order merging was less than ideal.  We therefore
carried out a new and independent reduction using a pipeline written by
Dan
Kelson\footnote{http://code.obs.carnegiescience.edu/carnegie-python-distribution}.
As part of these reductions, the blaze function was mostly removed by
dividing the spectra by the fitting functions used to normalize the
flat-field exposures.  To merge the echelle orders, the spectrum in each
order was normalized using cubic splines and the low signal-to-noise
ends were cut so that there was 5 to 10~\AA\ of overlap between adjacent
orders.
The signal-to-noise ratio per pixel near the peaks of the blaze function
was generally
in the range of $\approx 20-40$ for most of the spectra.

\subsection{UVES Spectroscopic Observations}

We also made use of five spectra obtained with the UVES instrument on
the second 8.2m telescope at the European Southern Observatory,
Paranal.  The observations were taken in service mode between 2004
December 19 and 2005 March 20.  The integration times were 2120
seconds each with seeing conditions between 0.8 and 2.2 arcseconds.  A
slit width of 2 arcseconds was used in combination with the dichroic
\#2 yielding a resolving power of $R=20,000$ and a useful wavelength
coverage of 3756~\AA\ to 4975~\AA.  The setup procedure for the instrument
ensures the star is precisely centered on the slit, even when the seeing
is less than the slit width used.
The spectra were fully
reprocessed, calibrated, and merged by the ESO UVES
pipeline\footnote{version 3.2, see
  http://www.eso.org/observing/dfo/quality/ reproUVES/processing.html}.
The typical signal-to-noise ratios per pixel were around 30 at a
wavelength of 4250~\AA.  Results based on these data have been
published by Val Baker et al.\ (2005).

\section{Photometric Observations}

\subsection{SMARTS Observations}

LMC X-3 was observed using the ANDICAM instrument on the SMARTS 1.3m
telescope at Cerro Tololo between 2007 October 5 and 2011 January 25.
The source was observed on most nights when it was available at an
airmass of less than $\approx 1.6$.  An observing sequence consisted
of observations in the $B$, $V$, $I$, and $J$ filters, with exposure
times of 180 seconds each in the optical filters and 30 seconds in
each of 15 dithered images in $J$.  The data were processed using the
SMARTS pipeline in IRAF.  Differential light curves were obtained
using two nearby stars in $B$, 4 stars in $V$, 6 stars in $I$, and 6
stars in $J$. The observations were placed on the standard scales
using observation of stars from Landolt (1992) for the optical and
stars from Persson et al.\ (1998) for $J$.  In all, observations of
standards from 45 nights in $B$, 44 nights in $V$, 46 nights in $I$,
and 76 nights in $J$ were used to establish the zero-points.

\subsection{Archival Data}

There have been previous studies of LMC X-3 in the optical, and we have
made use of published data from two sources.  Brocksopp et al.\ (2001)
carried out one of the larger studies, where they obtained optical
observations over a period spanning 6 years.  Specifically, observations
in $B$ and $V$ were made between 1993 and 1999 during 16 separate
observing runs, each of one to four weeks in duration.  The calibrated light
curves were kindly sent to us by C. Brocksopp.  Our second source of
archival optical data is van der Klis et al.\ (1985) who obtained $B$
and $V$-band CCD photometry of LMC X-3 from 1983 November 15-20.  The
times and differential $B$ and $V$ magnitudes were taken from their
Table 1.

\section{Spectroscopic Analysis}

Many of the quantities needed for the dynamical model discussed below
can be measured from the spectra.  We used the software tools SYNSPEC
(version 49) and SYNPLOT (version 2.1) written by Ivan Hubeny (Hubeny,
Stefl, \& Harmanec 1985; Hubeny 1998) to generate model spectra
interpolated from the BSTAR2006 grid (Lanz \& Hubeny 2007) to provide
templates for the radial velocity extraction and to determine the values
of the effective temperature, gravity, and rotational velocity of the
companion star.  We discuss below our improved radial velocity curves
and revised ephemeris, and the spectroscopic parameters of the companion
star.

\begin{figure}[t]
\epsscale{1.00}
\plotone{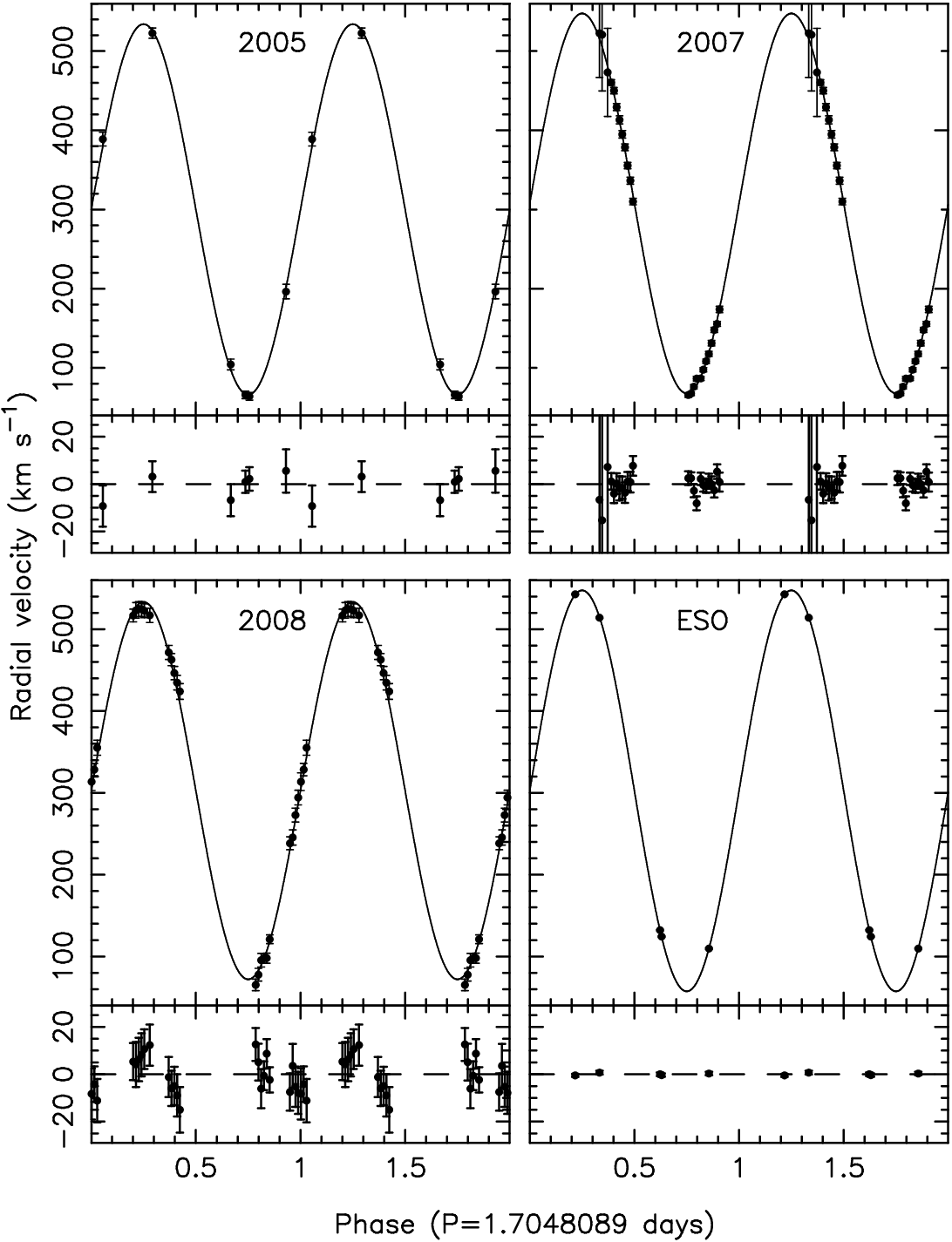}
\caption{The phased radial velocities of LMC X-3 from the different
observing runs are shown with the best-fitting models.  Two cycles are
shown for clarity.  See Table 1 for the model parameters.  }
\label{rvcomb}
\end{figure}

\begin{deluxetable}{rrrrr}
\tabletypesize{\footnotesize}
\tablecaption{Spectroscopic Parameters of LMC X-3}
\tablewidth{0pt}
\tablehead{
\colhead{parameter} &
\colhead{MIKE} &
\colhead{MIKE} &
\colhead{MIKE} &
\colhead{ESO} \\
\colhead{} &
\colhead{2005} &
\colhead{2007} &
\colhead{2008} &
\colhead{UVES} 
}
\startdata
period (days) & 1.7048089\tablenotemark{a} &1.7048089\tablenotemark{a} 
       &1.7048089\tablenotemark{a} &
1.7048089\tablenotemark{a}  \\
$T_0$ (HJD 2,450,000+) 
& $3391.1914\pm0.0069$ 
& $4454.9963\pm0.0024$ 
& $4523.1196\pm0.0032$ 
& $3449.1600\pm0.0007$  \\
$K_2$ (km s$^{-1}$) 
& $233.97\pm 3.46$
& $239.85\pm 2.19$
& $230.52\pm 2.91$
& $245.01\pm 0.42$  \\
$\gamma$ (km s$^{-1}$)  
& $300.08\pm 3.14$
& $307.72\pm 1.29$
& $302.62\pm 1.81$
& $303.43\pm 0.35$ \\
rms (km s$^{-1}$)  
& 5.472 
& 4.822
& 7.787
& 0.434 \\
$N$
& 6
& 24
& 24
& 5   \\
ASM intensity (counts s$^{-1}$)\tablenotemark{b} 
& 1.99, 0.05
& 0.86, 0.01
& 2.22, 0.02
& 2.37, 0.63  \\
$\llog$\tablenotemark{c} 
& $38.25\pm 0.13$
& $37.88\pm 0.42$ 
& $38.29\pm 0.08$
& $38.32\pm 0.18$  \\
$\Delta K$ (km s$^{-1}$)
& $3.97\pm 0.74$
& $2.26\pm 0.81$
& $4.28\pm 0.51$
& $4.47\pm 0.86$       \\
$K_{\rm corr}$ (km s$^{-1}$)
& $237.9 \pm 3.5$
& $242.1 \pm 2.3$
& $234.8 \pm 3.0$ 
& $249.5 \pm 1.0$
\enddata
\tablenotetext{a}{Fixed.}
\tablenotetext{b}{X-ray intensity from the {\em RXTE} 
All Sky Monitor. The quoted numbers are the average count rate and the standard
deviation of the individual measurements given in Table 2.}
\tablenotetext{c}{$L_{\rm x}$ (erg s$^{-1}$) = (ASM rate)$\times$($8.85\times 10^{37}$). The uncertainty 
includes the uncertainty on the individual ASM measurements.}
\label{tab1}
\end{deluxetable}

\subsection{Radial Velocities Measurements}

An improved ephemeris derived from the MIKE spectra and the spectra of
Cowley et al.\ (1983) was given in Song et al.\ (2010), and the reader
is referred to that publication for the full details.  The radial
velocities we derived from the re-reduced spectra confirm this
ephemeris.

An improved cross-correlation analysis (Tonry \& Davis 1979) was used to
derive radial velocities from all of the MIKE spectra and the UVES
spectra.  For the template we used a synthetic spectrum with the same
resolving power, wavelength sampling, and rotational velocity as the
object spectra.  The wavelength region used in the analysis was
$4000-4060$~\AA, $4158-4290$~\AA, $4376-4790$~\AA, and $4910-5000$~\AA,
which excludes the broad Balmer lines.  
Our previous ephemeris was found by using the MIKE radial velocities along
with the radial velocities from Cowley et al.\ (1983).  To account for
possible differences in the zero-points of the velocity scales, a
circular orbit model was fitted to each data set and the respective
systemic velocities were removed.  We repeated this exercise 
by using our new MIKE radial velocities, the UVES radial velocities,
and the Cowley et al.\ radial velocities and found
essentially the same ephemeris that was reported in Song et al.\ (2010).
We therefore adopt that ephemeris in the analysis reported below.

\begin{figure}[t]
\epsscale{1.00}
\plotone{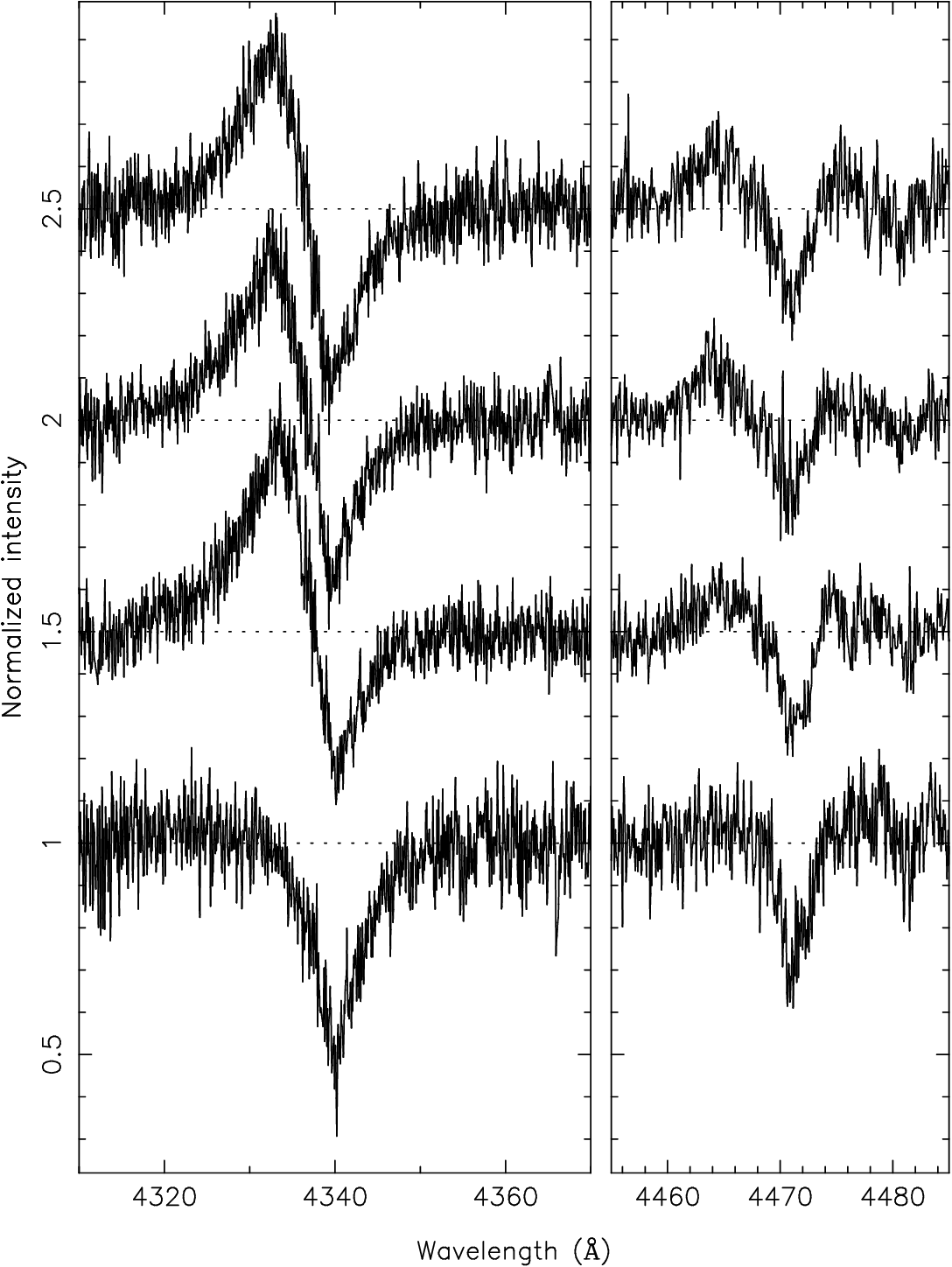}
\caption{Normalized spectra near H$\gamma$ (left) and He I (right) for
the first four MIKE observations from 2007 December 21, with the
earliest observation shown at the top.  The orbital phases are
0.333, 0.346, 0.372, and 0.389.
A clear emission component is
seen in H$\gamma$ and other Balmer lines.  While the signal-to-noise
ratios are relatively poor, this emission feature is relatively weak or
absent in the He lines.  }
\label{showBalmer}
\end{figure}

LMC X-3 is highly variable in X-rays. 
Given the
X-ray variability, the radial velocity measurements were divided up into
four sets: the measurements from the UVES spectra and the measurements
from the 2005, 2007, and 2008 MIKE spectra.  In Table 1 we give
average 
2-12 keV X-ray intensities from the {\em RXTE} All Sky Monitor (ASM).
Since the ASM did not continuously observe a given source,
some interpolation is needed.
We  used a smoothing and interpolation scheme based on both
the ASM and PCA data
to estimate the source brightness.  
The error bars given in the table come from a
Monte-Carlo code that varied the 
intensities using the nominal ASM and PCA error bars.
{ While not ideal, we believe this is a 
reasonable way to gauge the instantaneous X-ray intensity and its uncertainty. 
Each 2--12 keV intensity value was then converted to a 
bolometric X-ray luminosity using the 
procedure outlined in Appendix A.}
Finally, a three parameter sinusoid
was fitted to each data set with the period fixed at $P=1.7048089$
days
(the formal uncertainty on the
period is $0.0000011$ days, Song et al.\ 2010), 
yielding the parameters given in Table \ref{tab1}.  The phased
data and best fitting models are shown in Figure \ref{rvcomb}.  

\begin{figure}[h!]
\epsscale{1.00}
\plotone{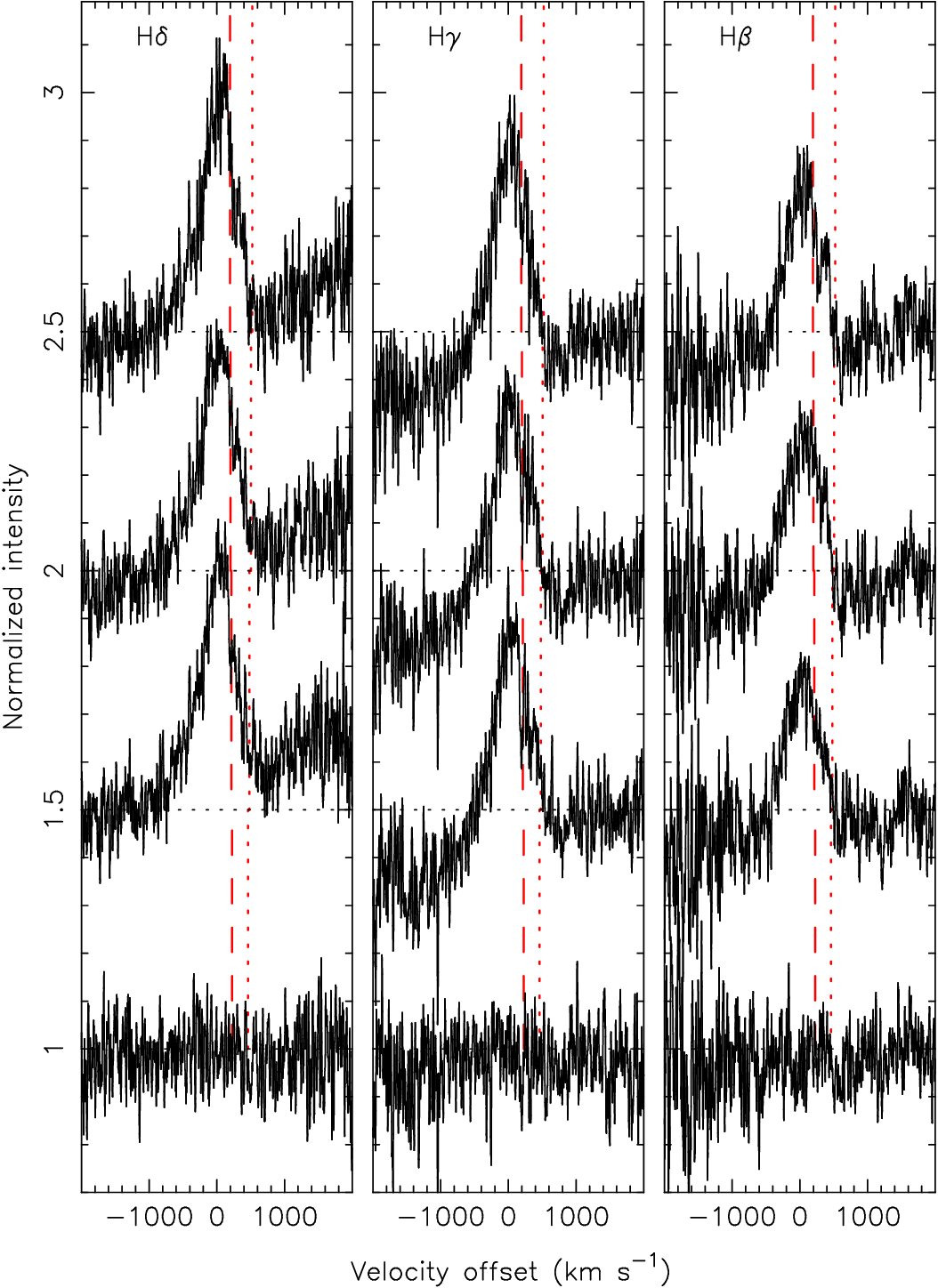}
\caption{This figure shows the differences between the first four
spectra from the night of 2007 December 21 relative to the fifth
spectrum from that night, in velocity units near H$\beta$, H$\gamma$,
and H$\delta$.  The orbital phases are
0.333, 0.346, 0.372, and 0.389.
The vertical dotted lines denote the radial velocity of
the secondary star, and the vertical dashed lines denote the approximate
radial velocity of the black hole.  The residual Balmer emission lines
are evident in the first three observations.  The features are
reasonably well fit by Gaussians with FWHMs near 600 km s$^{-1}$,
centroid velocities between 0 and 50 km s$^{-1}$ (for reference the
systemic velocity is 300 km s$^{-1}$), and equivalent widths between 3
and 4~\AA.  The emission lines appear to be completely gone by the fourth
observation.  }
\label{showemission}
\end{figure}

We find some scatter in the resulting $K$-velocities with a low value of
$K_2=230.52\pm 2.91$ km s$^{-1}$ for the 2008 MIKE spectra to a high value
of $245.01 \pm 0.42$ km s$^{-1}$ for the UVES spectra.  For comparison,
Cowley et al.\ (1983) found $K_2=235\pm 11$ km s$^{-1}$, and Val Baker et
al.\ (2007) found $K_2=242.4\pm 4.3$ km s$^{-1}$ from the UVES data (this
is their measurement before they applied a correction for X-ray
heating).  Also, the quality of the fits (judging by the rms of the
residuals) varies, ranging from 0.434 km s$^{-1}$ for the five ESO
measurements to 7.787 km s$^{-1}$ for the 24 MIKE measurements from
2008.  
For the three MIKE data sets, there is a weak inverse correlation between the 
$K$-velocity and the X-ray luminosity $L_{\rm x}$, which is discussed in \S5.2.

\subsection{Emission Lines}

Generally speaking, the optical spectra of LMC X-3 are dominated by the
absorption line component due to the secondary star.  However, during
the first three MIKE observations  from the night of 2007 December
21 at an orbital phase of 0.34, 
an emission feature was seen in the blue wings of the Balmer lines
(Figure \ref{showBalmer}).
A simple differencing process using the first five observations from
that night was used to isolate the emission components in the Balmer
lines.  We used the spectra directly from the reduction pipeline (counts
vs.\ wavelength for individual orders) with the blaze function mostly
removed.  The orders that contain H$\beta$, H$\gamma$, and H$\delta$
were each normalized to approximately unity by dividing by their
respective mean count rate.  The fifth spectrum was subtracted from each
of the first four spectra (the fourth spectrum has no emission lines and
serves as a ``control'' spectrum), and unity was added to the difference
spectra to put the continuum at 1.0. The results are shown in Figure
\ref{showemission}.  The difference spectra from the first three
observations show emission lines that are reasonably well modeled by
Gaussians with full widths at half maximum of $\approx 600$ km s$^{-1}$
and equivalent widths between 3 and 4~\AA.  The emission appears to be
gone by the fourth observation.  The He I lines show a hint of a
blue-shifted emission, albeit at a much smaller level than in the Balmer
lines.  The radial velocities derived using these three spectra (derived
using the He I and other metal lines) show more scatter than the other
measurements from 2007, but they are still within about 20 km s$^{-1}$
of the model curve (Figure \ref{rvcomb}).  The uncertainties on these
three points were inflated by a factor of 10 so that they would receive
very little weight in the sinusoid fit.

The emission features in the Balmer lines were observed in $\approx 6\%$
of the MIKE spectra, and in none of the UVES spectra.  Cowley et al.\
(1983) noted the ``occasional, very weak P Cygni emission at H$\beta$'',
but did not specify the number of spectra that contained this
feature.  The origin of these emission lines is not immediately clear.
The radial velocities of the emission lines from the first three 
observations are roughly consistent with the expected radial velocity 
of the black hole, which may suggest an origin in the accretion disk.  
However, if this is the case, the emitting region probably did not 
cover the entire disk because the profiles are not double-peaked.
 The rapid disappearance of the
emission may suggest an association with a rapidly varying region such
as the mass transfer stream.  However, if this is the case, the radial
velocity of the emitting gas should more closely match the radial
velocity of the secondary star (e.g.\ Gies \& Bolton 1986).

\begin{figure}[t!]
\epsscale{1.}
\plotone{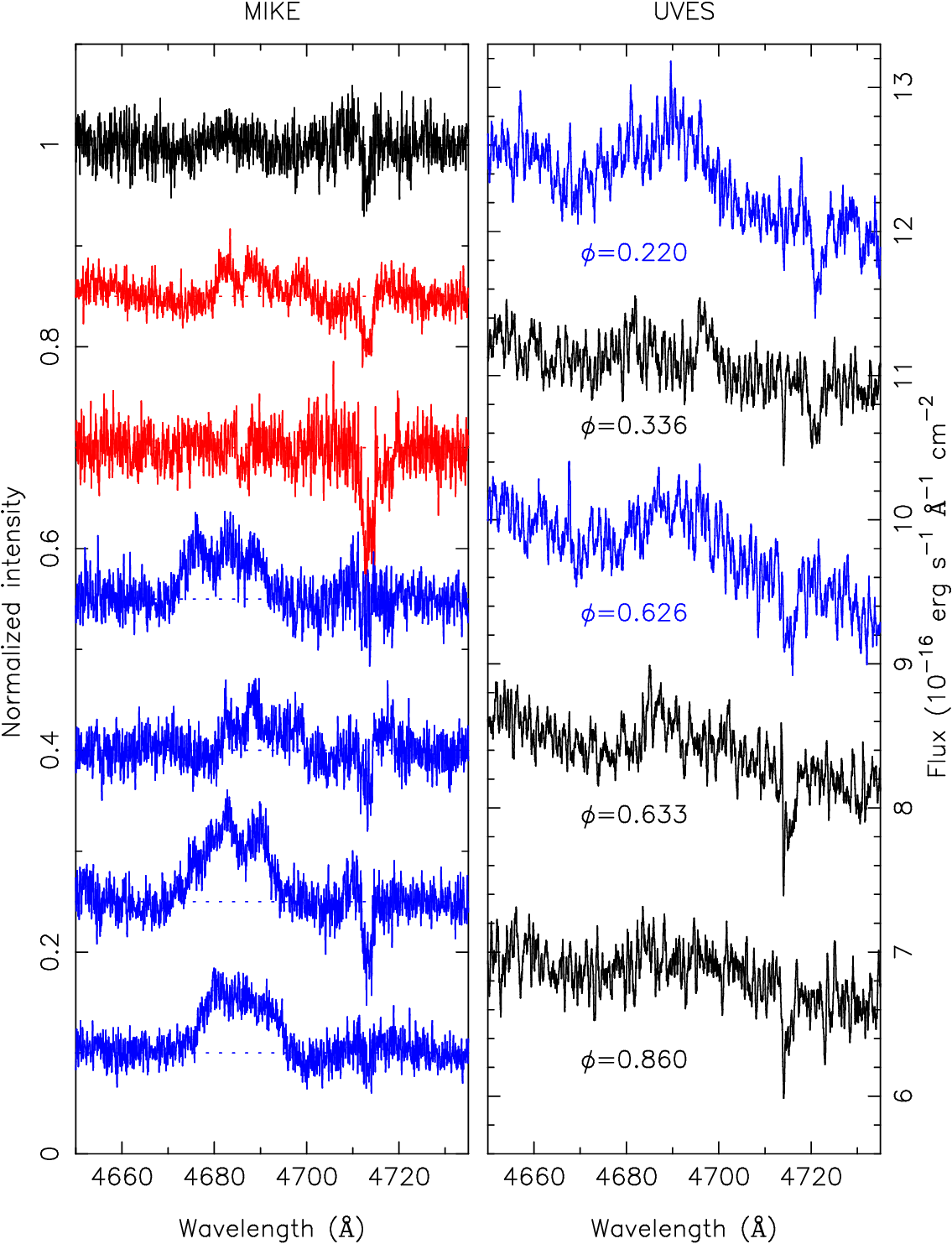}
\caption{Left: Normalized ``restframe'' spectra of the MIKE
observations near the He II $\lambda4686$
line, averaged from each night or run.  From top to bottom, the spectra
are from 2005 (6 nights), night 1 of 2007 (12 spectra), night 2 of 2007
(9 spectra), night 1 of 2008 (6 spectra), night 2 of 2008 (6 spectra),
night 3 of 2008 (5 spectra), and night 4 of 2008 (7 spectra).  No
emission feature is seen in the 2005 observations or in the spectrum
from the second night of 2007.  A broad emission line is present in all
of the spectra from 2008, especially from the last two nights.  The
absorption line near 4713~\AA\ is He I.
Right:  The five UVES observations are shown smoothed with 15 point
running means.  The vertical scale is in flux units, and vertical
offsets have been applied for clarity.  The spectra are ordered by
orbital phase, and no Doppler correction has been applied.  The two
spectra at orbital phases 0.220 and 0.626 (shown in blue) appear
to have the strongest He II emission line.}
\label{showNIII}
\end{figure}

In addition to the Balmer emission, there is, on occasion, a broad
emission feature which is almost certainly due to He II $\lambda4686$
(see Figure \ref{showNIII}).  This feature, which is commonly seen in
actively accreting X-ray binaries, seems to be strongest in the 2008
MIKE observations, and is either very weak or absent in the 2005 and
2007 MIKE observations. Perhaps not surprisingly, the mean X-ray
intensity was higher for the 2008 observations than it was for the 2005
and 2007 observations.  Owing to the large width (about 15~\AA),
properly normalizing the spectra to the local continuum is a challenge.
The MIKE specra are not flux calibrated so we cannot measure the
emission line flux.
Typical values for the equivalent 
widths of the He II feature seen in 2008 range from about 0.8 to 1.2~\AA.

The He II emission line is  detectable in two of
the five UVES spectra (Figure \ref{showNIII} shows
the UVES spectra smoothed with a 15 point running mean).  
The line flux is on the order of
$1.2\times 10^{-16}$ erg s$^{-1}$ cm$^{-2}$ \AA$^{-1}$,
and the equivalent width is about 1.4~\AA.
The two spectra where the line is detected were taken
2005 March 19 and 20 (at orbital phases 0.626
and 0.220).  The X-ray luminosity was about a factor of
1.5 times larger than it was during the three previous
UVES observations.  Although the signal-to-noise
ratios are not large, it does appear that the He II
emission line is moving out of phase relative to the 
nearby He I absorption line.  This suggests that
the He II emission is coming from the accretion disk and
not the heated part of the secondary star.

\subsection{Stellar Parameters and the Rotational Velocity}

\begin{figure*}[t!]
\includegraphics[scale=0.7,angle=-90]{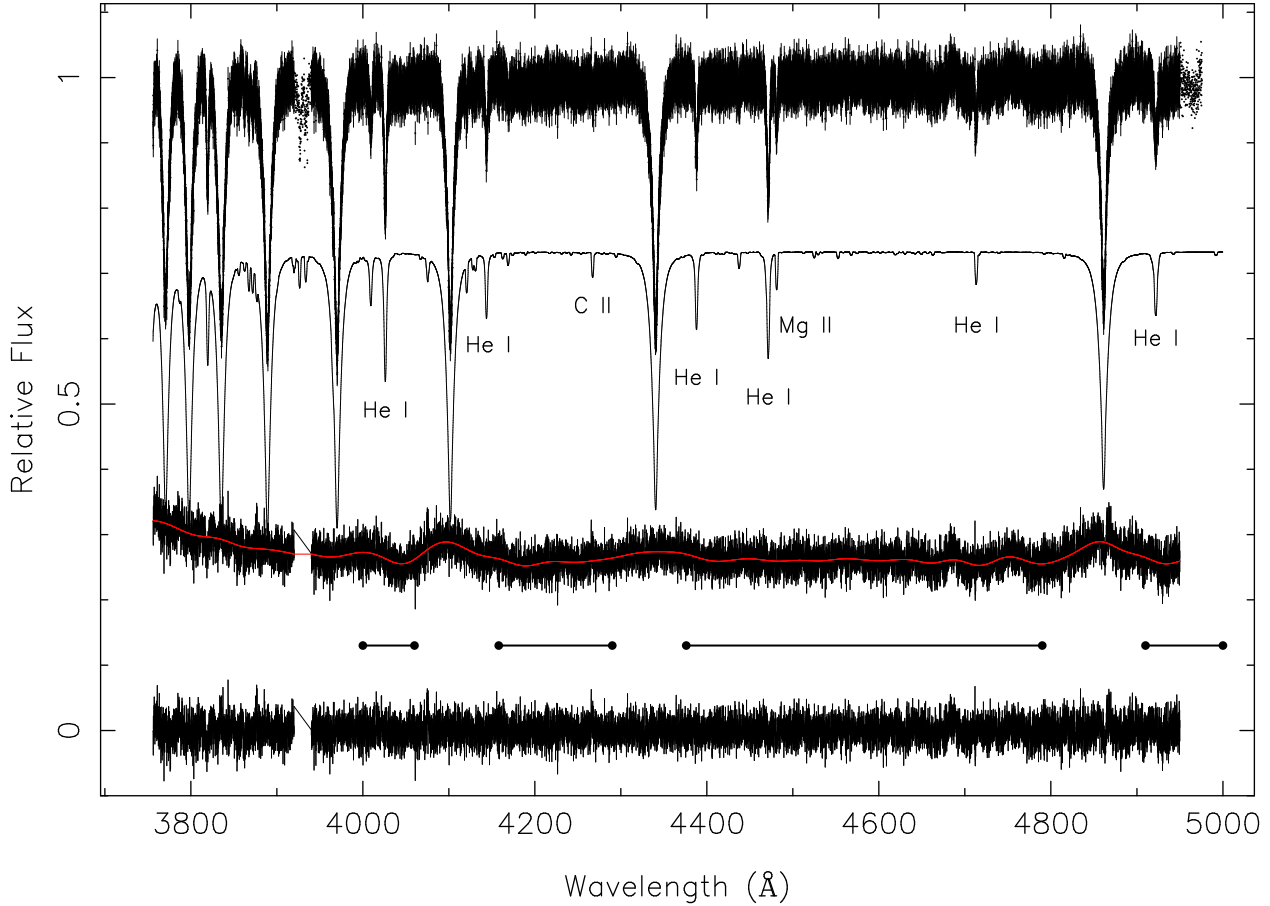}
\caption{An illustration of the process used to model the spectra is
shown.  The spectra are, from top to bottom, the normalized restframe
spectrum made from the UVES spectra; a normalized model spectrum with
$T_{\rm eff}=16,435$, $\log g=3.801$, and $V_{\rm rot}\sin i=121.4$ km
s$^{-1}$, scaled by 0.76; the difference between the restframe spectrum
and the scaled model; and the difference between the residuals and a
smoothed version of itself.  The stellar features are removed in the
final spectrum.  In this case, the star contributes 76\% of the light in
this wavelength region.  }
\label{weightproc}
\end{figure*}

As noted above, LMC X-3 is extremely variable in the optical, where
excursions of up to $\approx 1$ mag have been observed (Brocksopp et
al.\ 2001).  
Some B-stars are known to be pulsational variables (e.g.\ stars of the 
$\beta$ Cephei type) with typical periods less than 
about one day.  Both the light and radial velocity curves 
can be affected by these pulsations.
Since many of the light curves of LMC X-3 look 
cleanly ellipsoidal and since the well-sampled 
radial velocity curves are smoothly sinusoidal, we will assume that 
(apart from ellipsoidal variability) the B-star in LMC X-3 is intrinsically 
non-variable.  Given this, the large excursions in the mean 
brightness level are most likely due to a variable component of disk 
light, which implies that the spectra contain at least a 
small contribution of light from the accretion disk.
We therefore used the method outlined in Marsh, Robinson, \& Wood
(1994) to decompose the spectra into the stellar absorption line
component and the accretion disk component.  Briefly, the spectrum to be
fitted is normalized to its continuum.  A model spectrum in constructed
from the BSTAR2006 models (using the LMC metallicity grid) with a given
temperature $T_{\rm eff}$, gravity $\log g$, and rotational velocity
$V_{\rm rot}\sin i$.  The model is scaled by a weight $k$ 
(where $k$ is the fraction of the total light that is contributed by the star)
and subtracted
from the observed spectrum.  A heavily smoothed version of the
difference spectrum is subtracted from the difference spectrum itself,
and the $\chi^2$ of the residuals is computed.  The value of $k$ is
varied until the value of $\chi^2$ is minimized.  Spectra at various
stages of this process are shown in Figure \ref{weightproc}.

\begin{figure}[t!]
\includegraphics[scale=0.34,angle=-90]{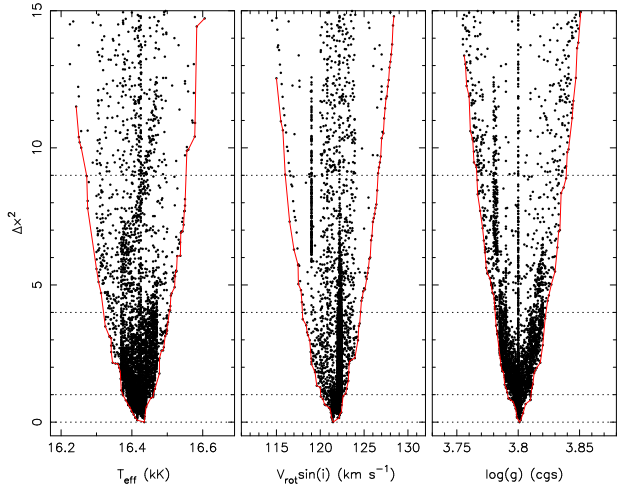}
\caption{Plots of $\chi^2$ vs.\ the effective temperature
(left), the rotational velocity (middle), and the
gravity (right) are shown from the fit of the average UVES
restframe spectrum. 
The horizontal dotted lines
denote the $1\sigma$, $2\sigma$, and $3\sigma$ confidence regions.
}
\label{plotchi}
\end{figure}

\begin{figure}[t!]
\includegraphics[scale=0.34,angle=-90]{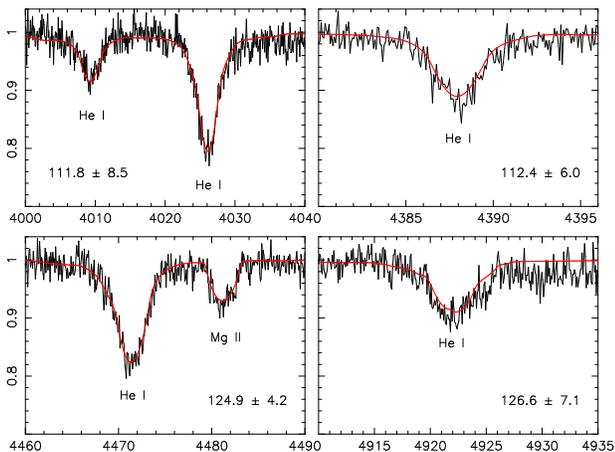}
\caption{Model fits to individual lines (or close pairs) 
in the restframe UVES spectrum.
The derived rotational velocity
is shown in each panel.
}
\label{plotHEfits}
\end{figure}

For the purposes of determining the rotational velocity (which is needed
for the dynamical model), the UVES spectra are superior to the MIKE
spectra owing to their better signal-to-noise, flux calibration, and order
merging.  We tried three ways of fitting the spectra to derive a
rotational velocity:  (i) using all H, He, and metallic lines in
the (Doppler-corrected) 
averaged UVES spectrum; (ii) using all H, He, and metallic
lines in each UVES Doppler-corrected
spectrum and averaging the individual
measurements; and (iii) using 
individual strong He and metallic lines
in the average UVES spectrum.  The fourth possible way, namely using
individual strong He and metallic lines in each spectrum did not
add any useful information owing to the relatively low signal-to-noise.
For each case, the decomposition technique was applied using 
template spectra drawn from a dense model grid
with a wide range of temperatures, gravities, and rotational
velocities.  The $\chi^2$ values for each template were recorded, and
marginalized distributions of $\chi^2$ over each parameter of interest
were generated (see Figure \ref{plotchi} for the distributions from the
fit to the average UVES spectrum).  The parameter values are summarized
in Table \ref{BAT}.  The rotational velocity derived from the average
spectrum (e.g.\ case (i) above)
is $V_{\rm rot}\sin i=121.4\pm 1.4$ km s$^{-1}$.  For case
(ii), the 
average of the rotational velocities derived from the individual spectra
is $V_{\rm rot}\sin i=118.5\pm 6.6$ km s$^{-1}$, where the stated
uncertainty is the formal error in the  mean.
Model fits to 
individual (or close pairs) of
He I lines in the average UVES spectrum (case (iii) above)
are shown in
Figure \ref{plotHEfits}.  The  average of the four
measurements is $ 118.9\pm 3.6$ km s$^{-1}$.

\begin{figure*}[t!]
\includegraphics[scale=0.67,angle=-90]{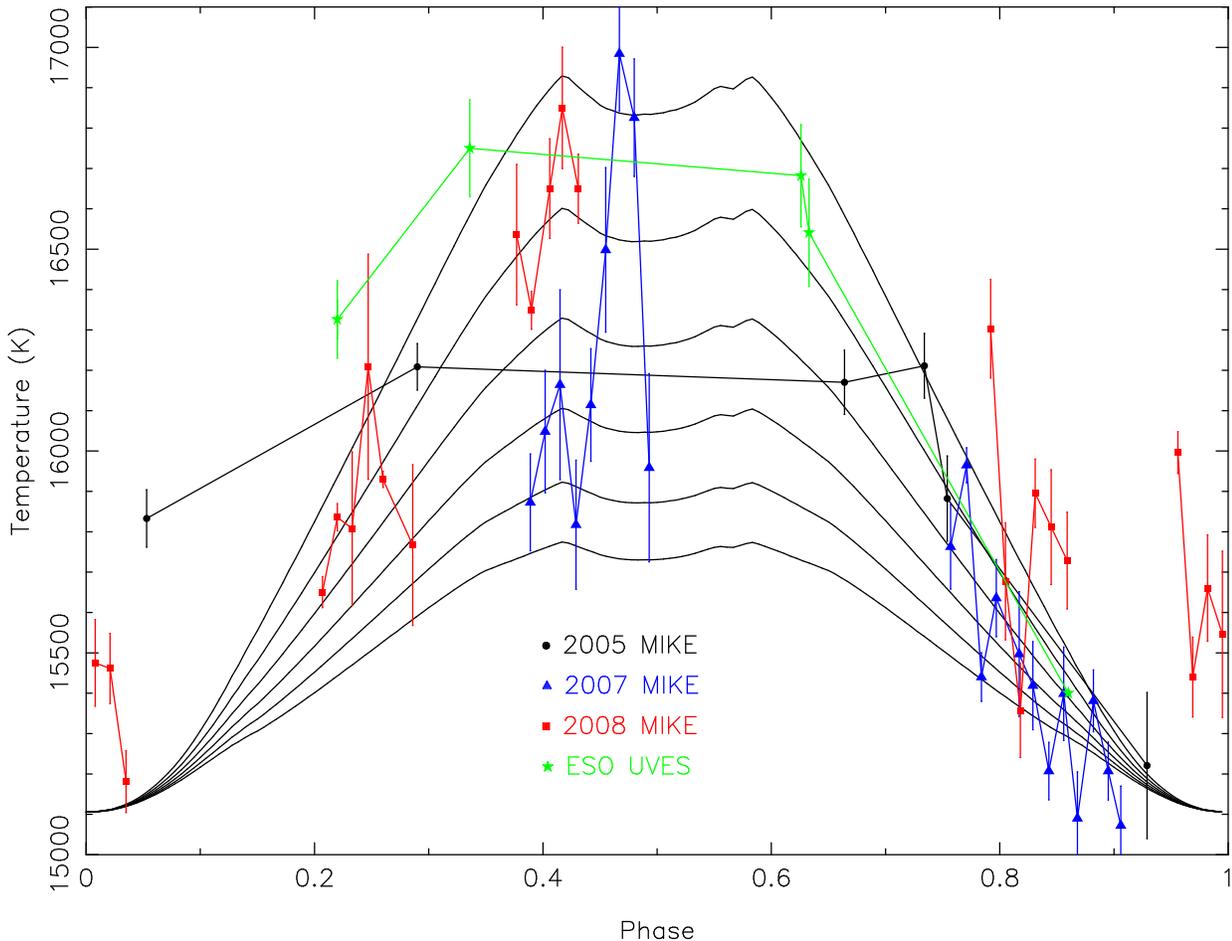}
\caption{The effective temperatures derived from the individual spectra
are shown as a function of the orbital phase.  Measurements from the same
night or run (in the case of the 2005 MIKE spectra and the
ESO UVES spectra) are connected by lines.
The solid curves are models with an intrinsic stellar
temperature of 15,100 K (chosen to
illustrate the effect), and $\llog=38.0$ 
(lower-most curve) to
$\llog=38.5$ in equal intervals.
}
\label{temperaturetrend}
\end{figure*}

The rotational velocities derived for the three cases above are all in
good agreement.  
{We note that 
the X-ray luminosity during the first three UVES
observations
was $\approx 60\%$ lower than it was for the last two observations
(see Table
\ref{BAT}.)} 
Given that averaging spectra with different X-ray luminosities 
(as in cases (i) and (iii)) may introduce
small systematic uncertainties, we adopt the results for case (ii), namely
$V_{\rm rot}\sin i=118.5\pm 6.6$ km s$^{-1}$.  

For comparison, we also fitted the 51 individual MIKE spectra
that did not have emission lines
and the restframe spectra made from nightly averages (in the case of
the 2005 run all six spectra were combined into a restframe spectrum).
The results are given in Table \ref{BAT}.  The average of
the 51 measurements is $127.2 \pm 3.5$ km s$^{-1}$ and the
the median
value is 118.2 km s$^{-1}$, both of which agrees well with the rotational
velocity derived from the UVES spectra that we adopt above.

As a consistency check, we point out that
owing to the peculiarities of Roche geometry, the mean density of a
Roche-lobe filling star mainly depends on the orbital period of
the binary.  For a fixed orbital period, there is little variation in
the stellar density over a wide range of mass ratios (e.g., Eggleton
1983).  In that same vein, the surface gravity of the companion is
weakly dependent on the mass ratio of the binary and is usually well
determined from the light curve models.  In the model discussed below,
we find $\log g=3.73$ for the secondary star.  This dynamical
measurement of the gravity is in good agreement with the
spectroscopically-determined values: All of the gravities measured
from the individual UVES spectra are within 0.12 dex of the dynamical
measurement, and the gravity found from the mean UVES spectrum is
within 0.05 dex.  The gravities from the individual MIKE spectra show
considerably more scatter, where the mean gravity is 
$\log g=3.45$ and the standard deviation is 0.27 dex.

\begin{deluxetable}{rrrrrrr}
\addtocounter{table}{1}
\tabletypesize{\footnotesize}
\tablecaption{Photometric Data for LMC X-3}\label{nametab}
\tablewidth{0pt}
\tablehead{
\colhead{\#} &
\colhead{Data subset} &
\colhead{JD range} &
\colhead{UT range} &
\colhead{Number of} &
\colhead{Number of} &
\colhead{Reference\tablenotemark{a}}  \\
\colhead{} &
\colhead{label} &
\colhead{(-2,440,000)} &
\colhead{(YYYY/MM/DD)} &
\colhead{$V$-band} &
\colhead{$B$-band} &
\colhead{}}
\startdata
1 &A93  & 9266.86-9275.87   & 1993/10/06-1993/10/15 & 72 & 6  &  1 \\
2 &W94  & 9705.62-9726.58   & 1994/12/19-1995/01/09 & 24 & 23 &  1 \\
3 &W95  & 10059.55-10069.83 & 1995/12/08-1995/12/18 & 54 & 52 &  1 \\
4 &S96  & 10148.59-10156.59 & 1996/03/06-1996/03/14 & 12 & 11 &  1 \\
5 &A96a & 10324.84-10340.90 & 1996/08/29-1996/09/14 & 63 & 45 &  1 \\
6 &A96b & 10343.78-10353.89 & 1996/09/17-1996/09/27 & 42 & 44 &  1 \\
7 &W96a & 10413.58-10423.86 & 1996/11/26-1996/12/06 & 38 & 25 &  1 \\
8 &W96b & 10424.70-10441.86 & 1996/12/07-1996/12/24 & 47 & 39 &  1 \\
9 &S98  & 10856.72-10884.51 & 1998/02/12-1998/03/12 & 95 & 42 &  1 \\
10 &W98  & 11123.71-11142.86 & 1998/11/06-1998/11/25 & 73 & 26 &  1 \\
11 &vdK  & 5653.77-5658.81   & 1983/11/15-1983/11/20 & 25 & 10 &  2\\
12 &Halved & 15148.59-15587.68 & 2009/11/13-2011/01/26 & 1116 & 695 & 3 \\
13 &Quartered & 14822.72-15553.69 & 2008/08/13-2010/12/23 & 74 & 73 & 3 \\
\enddata
\tablenotetext{a}{1:  Brocksopp et al.\ (2001); 2: van der Klis et al.\ (1985);
3: this work}
\label{phototab}
\end{deluxetable}

We note that the temperatures measured for the individual spectra vary
with orbital phase, as shown in Figure \ref{temperaturetrend}.
Observations taken near phase 0 (inferior conjunction of the companion
star) tend to have smaller effective temperatures whereas observations
taken near phase 0.5 (when the companion star is farthest from the
observer) tend to have higher temperatures.  The change in the observed
temperature with orbital phase is almost certainly due to X-ray heating.
Near phase 0, the heated part of the star is mostly directed away from
the observer, so the parts of the star that are visible are cooler.  On
the other hand, the heated face of the star is most visible near phase
0.5, and as a result the observed temperature near that phase tends to
be higher.  Using the ellipsoidal model discussed below, we computed the
intensity-weighted average temperature of the visible parts of the star
as a function of the orbital phase for X-ray luminosities between
$10^{38}$ and $10^{38.5}$ ergs s$^{-1}$ (which bracket the typical
observed values); the results are shown in Figure
\ref{temperaturetrend}.  Although the scatter is large, these models
reproduce the changes seen in the temperature with orbital phase.  Near
phase 0, there is little variation in the observed temperature,
regardless of the X-ray luminosity.  On the other hand, the observed
temperature near phase 0.5 is sensitive to the value of the X-ray
luminosity.  We cannot expect a perfect match between these simple
models and the observations since the observations have large
uncertainties and since the X-ray luminosity can change significantly
on short time scales.  Based on the relatively small number of spectra
taken near phase 0, we adopt a range of $15,000 \le T_{\rm eff}\le
15,500$ K for the effective temperature the secondary would have if it
were unheated and spherical.

Finally, one can see in Table \ref{BAT} some trends in $k$, 
the fraction of the total light contributed by the star.
During 2007, when the average X-ray intensity was lower, the
stellar fractions are usually close to 0.9.  During 2008, when the
average X-ray intensity was higher, the stellar fractions are closer to
0.7.  Thus it seems that the amount of disk light (e.g.\ the value of
$1-k$) roughly tracks the observed X-ray intensity.  Cowley et al.\
(1983) also found diluted line strengths, noting that ``although the
{\em relative} line strengths are normal, all the lines are $\sim 2$
times too weak compared with a standard B3 main-sequence star...''.

\section{Dynamical Models}

\subsection{Light Curve Selection}

LMC X-3 is a highly variable source at both X-ray and optical
wavelengths.  Thus it is not advisable to simply model all of the
photometric data simultaneously.  We divided these data into 13 
data sets
as follows.  Brocksopp et al.\ (2001) had 16 different observing
sessions, each lasting about one to four weeks.  They labeled these
sessions after the season and year, as in S96 for the (Northern
Hemisphere) spring of 1996,
W98 for the (Northern Hemisphere)
winter of 1998, etc.  The folded light curves from eight of
these sessions had excessive scatter, too few points, or both, and were
discarded: A95, A97, A98, S95, S97, S99, W93, and W97.  In the case of
two extended sessions, A96 and W96, we subdivided each session into a
pair of sessions, thereby increasing our sample of clean folded light
curves.  Thus, the data of Brocksopp et al.\ (2001) yielded 10 sets of
light curves: A93, A96a, A96b, S96, S98, W94, W95, W96a, W96b, and W98.

The $B$ and $V$ light curves taken from van der Klis 
et al.\ (1985) constitute an eleventh data set.
There are 25 measurements in $V$ and 10
measurements in $B$, taken over a span of five nights.

In contrast to the Brocksopp et al.\ (2001) data, which consisted of
targeted observing runs over relatively short timescales, the SMARTS
observations consist of usually one set of $B$, $V$, $I$ and $J$
observations per night over the span of an entire season.  For this
sampling, the long-term aperiodic variability of the source is a
significant problem in modeling the ellipsoidal light curves.  We filter
out much of the variability by using the {\it RXTE} All-Sky Monitor
(ASM) X-ray light curve, which we interpolate after smoothing it using a
Gaussian kernel with a width of one week.  As a minor point, we reject
OIR data that deviate by more than $2.5\sigma$ from the local mean,
where $\sigma$ is the local rms; less than 1\% of the optical data and
$\sim 2.5\%$ of the $J$-band data were rejected.  In total, only 17 data
points out of 1900 were rejected.

\begin{deluxetable}{llll}
\tablecaption{ELC model parameters\label{freeparm}}
\tablewidth{0pt}
\tablehead{  
\colhead{Parameter} &
\colhead{Meaning} &
\colhead{Lower} & 
\colhead{Upper} \\
\colhead{} &
\colhead{} &
\colhead{bound} &
\colhead{bound} }
\startdata
$i$  & inclination (deg) & 40 &80 \\
$L_{\rm x}$ & X-ray luminosity (erg s$^{-1}$) & 36.0 &38.5 \\
$K_2$ & $K$-velocity (km s$^{-1}$) & 225 & 255 \\
$Q$ & mass ratio ($M_2/M_1$)  & 1.6 & 2.4  \\
$\Delta\phi$ & phase shift parameter & -0.04 & 0.04 \\
$T_{\rm disk}$ & temperature at inner disk edge (K) & 15,000 & 50,000 \\
$r_{\rm out}$ & radius of outer accretion disk edge & 0.40& 0.99 \\
$\xi$ & power-lower exponent on disk temperature profile & -0.80 & 0.0 \\
 $\beta$ & opening angle of disk rim (deg) & 1.0 & 5.0\\
$s_{\rm spot}$ & temperature factor for disk spot & 0.9 & 9.8 \\
$\theta_{\rm spot}$ & longitude of disk spot (deg) & 0.0 & 360.0 \\
$r_{\rm cut}$ & cut-off radius of disk spot & 0.50 & 0.99 \\
$w_{\rm spot}$ & angular width of disk spot (deg) & 1.0 & 50.0
\enddata
\end{deluxetable}
\addtocounter{table}{-1}

We consider two selections of the SMARTS data.  For the first of these,
which we refer to as ``X-quartered,'' we use the smoothed X-ray light
curve in two ways to filter the OIR data.  (1) We select only those OIR
data for which the simultaneous X-ray intensity is among the faintest
25\% of the ASM data record, when X-ray heating is minimal and the
stellar component of light is most dominant.  The upper flux limit to
this selection corresponds to a threshold ASM count rate of 0.88 
counts per second
(equivalent to $L \approx 20\% L_{\rm Edd}$).  (2) In Steiner et
al. (2013), we demonstrate that the X-ray emission lags the optical
emission by $\sim 15$ days.  Therefore, we reject any OIR data for which
the X-ray emission 15 days earlier is above the same threshold of 0.88
ASM counts per second.  
This criterion eliminates data taken during times when the
disk is brightest.  These two cuts, which serve to substantially reduce
the OIR variability induced by accretion, are stringent; only 285/1883
(16\%) of the OIR data survive.

We refer to the second set of selected data as ``X-halved.''  In this
case, we select the OIR data for which the simultaneous X-ray intensity
is one-half or less the average intensity (which corresponds to an ASM
count rate of 1.45 counts per second).  
However, unlike the X-quartered case, we
here employ the model described in Steiner et al.\ (2013) to compute the
OIR emission attributable to X-ray heating and to viscous dissipation in
the outer disk.  For the selected data, we subtract from the observed
OIR fluxes in each band the contributions predicted by the model due to
the disk emission and to X-ray reprocessing.  Figure 7 (and related
text) in Steiner et al.\ (2013) makes clear the efficacy of this approach
to filtering the data.  On average, the fluxes in each band are
corrected downward by $\sim10-15\%$.  The total number of selected data
points is 1013 out of 1883 (54\%).

Table \ref{phototab} gives a summary of the photometric data
sets used, including the names, the date ranges, and the number of
observations in $B$ and $V$ available for each set.

\subsection{ELC Model}

\subsubsection{Basic Setup and Light Curve Fits}

We used the ELC code (Orosz \& Hauschildt 2000) to construct our
dynamical model of LMC X-3.  There are two sources of optical light: the
secondary star and the accretion disk.  We assume the secondary star is
in a circular orbit with synchronous rotation, and that it fills its
Roche lobe.  The models are insensitive to the temperature of the
secondary star, and we use $T_{\rm eff}=15,500$ K and consequently set
the gravity darkening exponent to 0.25.  Specific intensities were
computed from model atmospheres from the BSTAR2006 grid (Lanz \& Hubeny
2007) with the LMC metallicity.  The reprocessing of X-ray radiation
from near the compact object
(hereafter ``X-ray heating'') can alter the
temperatures of parts of the secondary star. We use a simple model
based on Wilson (1990, see also Zhang,
Robinson, \& Nather 1986) to account for the heating.  
The overall amount of X-ray heating is
controlled by the X-ray luminosity $\log L_{\rm x}$. 
Let
$F_{\rm irr}$ be the incident flux of X-rays that can be seen
from a given point on the secondary star with coordinates 
($x$,$y$,$z$).  If the reprocessed light is completely thermalized,
then the new temperature of the point on the secondary becomes
$$
T_{\rm new}(x,y,z)^4=T_{\rm old}(x,y,z)^4+
\alpha F_{\rm irr}/\sigma
$$
where $\sigma$ is the Stefan-Boltzmann constant. The normal range of the
parameter $\alpha$ is 0.0-1.0: 
when the $\alpha=1$ 
the reprocessed X-ray radiation is immediately reradiated.  For 
our models of LMC X-3, 
the X-ray emitting
area was assumed to have a disk geometry, and 
we used
$\alpha=1.0$.  
The basic accretion disk
has four parameters, namely its outer radius $r_{\rm out}$ (expressed as
a fraction of the black hole's Roche lobe), 
the opening angle of the outer
disk
$\beta_{\rm rim}$, the temperature at the inner edge $T_{\rm disk}$, and
the power-law exponent on the temperature profile $\xi$.  
After some experimentation, it was found 
that adding a hot spot to the rim of the accretion disk 
improved the fits by allowing the disk 
light to be modulated with phase.  The 4 parameters needed to specify 
the spot on the accretion disk are $\theta_{\rm spot}$ (its 
longitude 
relative to the line connecting the two stars), 
$w_{\rm spot}$
(its angular 
width), $r_{\rm cut}$
(its radial extent), and $s_{\rm spot}$ (its temperature factor
by which the underlying temperature in a spot region is scaled).  
Note that the accretion disk can shield parts of the star from the effects of 
X-ray heating, an effect that is completely accounted for in the model.  
X-ray heating of the disk itself is accommodated by varying the 
temperature profile of the disk.  To complete the model,
the scale of the binary must be specified, and
here we used the inclination $i$, the mass ratio $Q\equiv M_{\rm
BH}/M_2$, and the $K$-velocity of the secondary star.  
All of the light curves were phased on the adopted ephemeris, and a
phase shift parameter $\Delta\phi$
was included to account for the small uncertainty
in the ephemeris.

We have additional observational constraints on the model that are not
directly tied to the optical light curves.  (1) 
The absence of an X-ray eclipse puts an upper limit on the
inclination of the binary (this upper limit depends on the mass ratio).
This constraint is imposed by giving models that produce an X-ray
eclipse a very large $\chi^2$ value.  (2)
As noted earlier, we have measured
the projected rotational velocity of the secondary star $V_{\rm rot}\sin
i$, and this puts a constraint on the mass ratio (e.g.\ Wade \& Horne
1988).  For a given set of model parameters, the expected value of the
rotational velocity is computed and the $\chi^2$ contribution relative
to the observed value is computed and added to the total $\chi^2$.  
(3) The
measured $K$-velocity of the secondary star constrains the binary scale,
and its contribution to the $\chi^2$ is computed in a similar fashion as
the contribution of the rotational velocity.  (Note the $K$-velocity is
also used as an input parameter and it is sampled from a uniform
distribution by the optimizing codes discussed below.)  
Finally, (4) the disk
fractions measured from the spectra limit the amount of parameter space
allowed.  Here we use a two part contribution to the $\chi^2$: $\chi^2_k=0$
if the computed value of $1-k$ for a model is less than 0.15, and
$\chi^2_k=[(k_{\rm mod}-0.15)/0.1]^2$ when $1-k$ is greater than 0.15,
where $k_{\rm mod}$ is the model disk fraction.  Models with very large
disk fractions are disfavored, whereas models with disk fractions near
the observed values of $\sim 0.1$ to 0.3 are given similar weights.

\begin{figure}[t!]
\epsscale{1.00}
\plotone{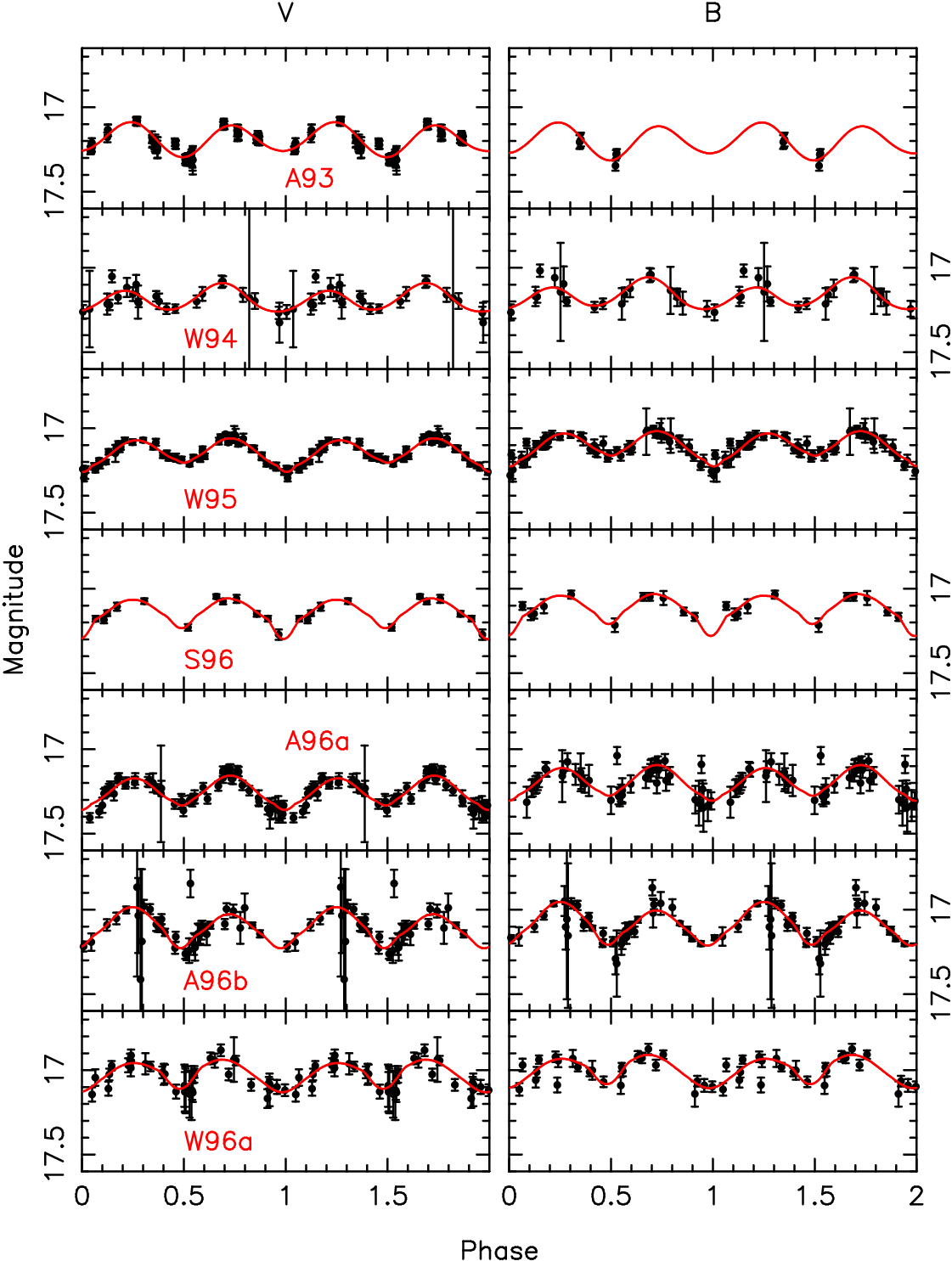}
\caption{The phased $V$ light curves (left panels) and
$B$ 
light curves (right panels) of LMC X-3 for 7 
of the 13 observing runs are shown along with the best-fitting models.
}
\label{lcfit1}
\end{figure}

\begin{figure}[h!]
\epsscale{1.}
\plotone{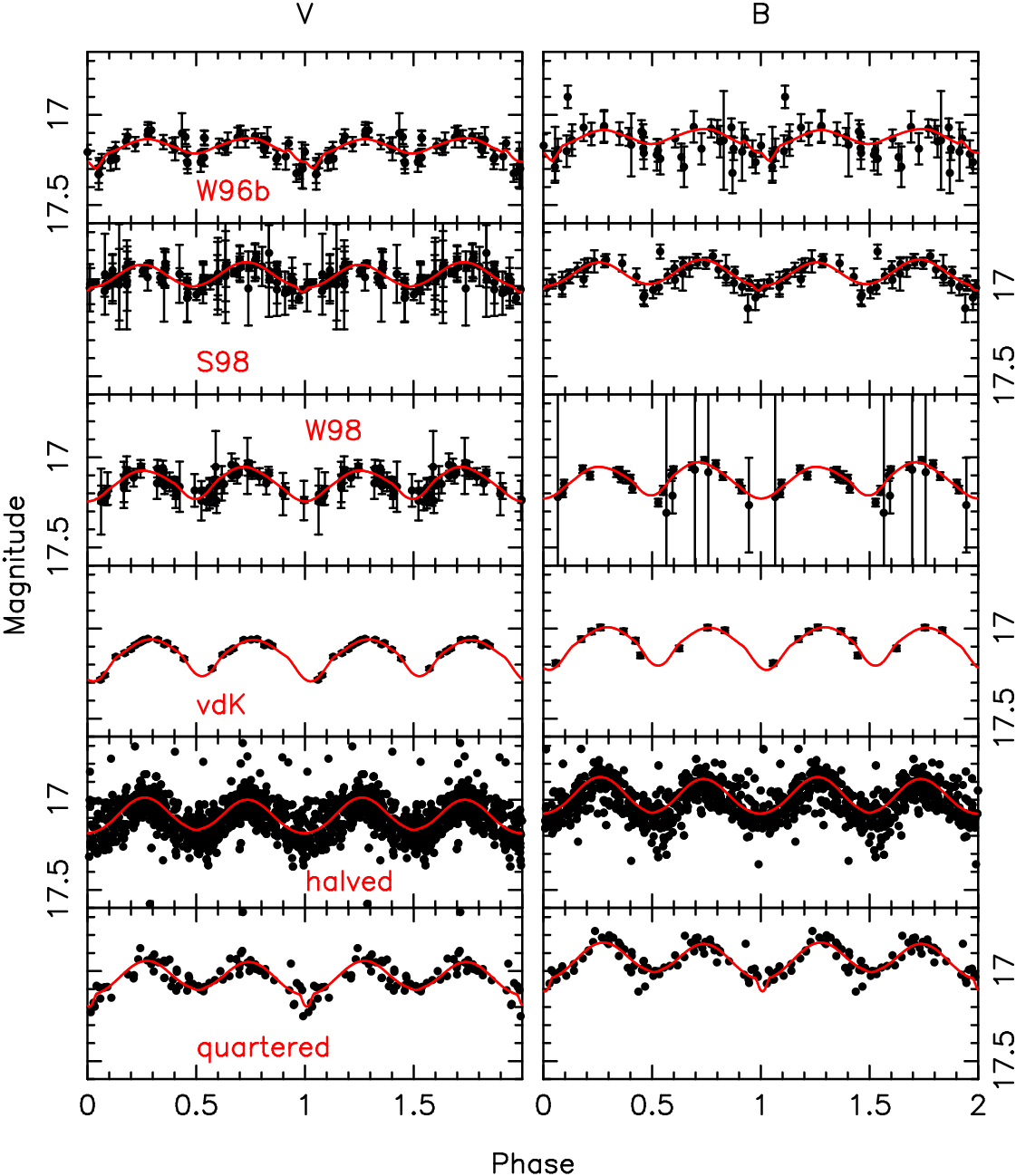}
\caption{The phased $V$ light curves (left panels) and
$B$ 
light curves (right panels) of LMC X-3 for the other 6 observing runs 
are shown along with the best-fitting models.
}
\label{lcfit2}
\end{figure}

\begin{figure}[t!]
\epsscale{1.}
\plotone{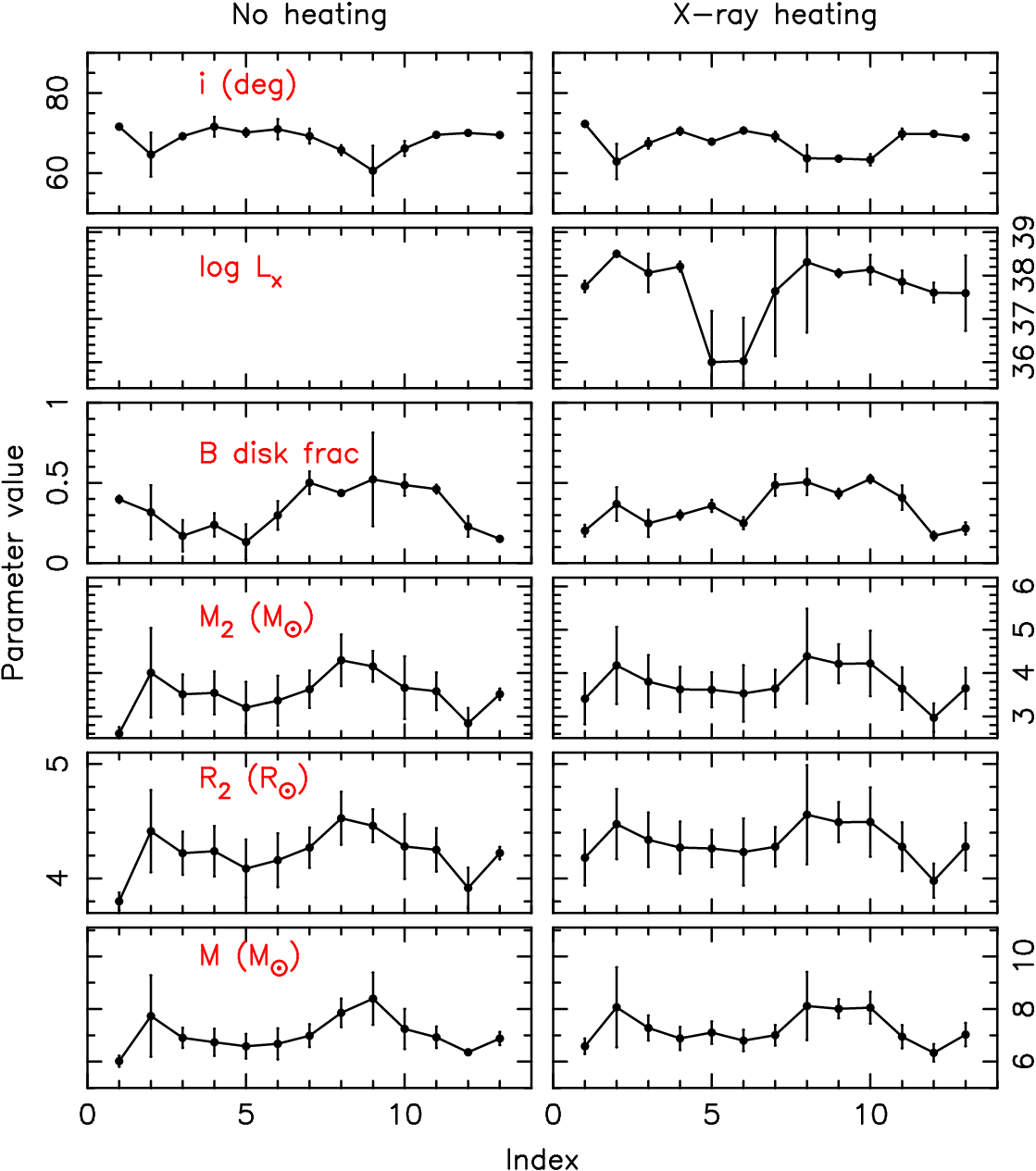}
\caption{The values of selected parameters from the ELC
fits plotted as a function of the season
(the order of the indices corresponds to the order the
light curves are plotted in Figures \protect\ref{lcfit1}
and \protect\ref{lcfit2}).  The left-hand
panels show the models with no X-ray heating, and the right-hand
panels show the models with X-ray heating.
}
\label{showresults}
\end{figure}

Our model has a total of 13 free parameters, which are summarized
in Table \ref{freeparm}.
The phased light curves were fitted using three of ELC's optimizers:
a Monte Carlo Markov Chain (see Tegmark 2004), a 
genetic algorithm (Charbonneau 1995), and a Differential Evolution
Monte Carlo Markov Chain (Ter Braak 2006).  The ranges of each of
the free parameters are given in Table \ref{freeparm}.  After a large
number of iterations for each technique, the model giving the smallest
$\chi^2$ was identified (Figures \ref{lcfit1} and
\ref{lcfit2}).  The uncertainties in the fitted and derived
parameters were found from the marginalized distributions of $\chi^2$.
The results are given in Tables \ref{tab3a}-\ref{tab3e}, and
graphically in Figure \ref{showresults}.  
As a check, we also fitted the phased light curves assuming no
X-ray heating, using only 12 free parameters with the same optimizing
schemes.     
In all of the 13 sets, 
the derived inclinations are within 11 degrees of each other 
and range from $62.9^{\circ}$ and $72.3^{\circ}$ for the models 
with X-ray heating and from $60.5^{\circ}$ and $71.6^{\circ}$
for the models without X-ray heating.
The other parameters (apart from the $B/V$ disk fractions) are likewise 
convergent for the 13 data sets.  We find that the models with and without X-ray heating 
generally return very similar parameter values.

Some of the model light curves have a small dip near phase
0.0.  
The dip is especially noticeable in the quartered light curve,
but it is present in other light curves such as W95.
This dip is caused by a grazing eclipse of
the outer rim of the accretion disk by the secondary star.  
Given the quality of the light curves, we cannot tell
for certain if such a grazing eclipse is real.  If it is real,
then the inclination would be tightly constrained as there is 
a very small ranges of inclinations were eclipses of the outer disk occur
but eclipses of the X-ray source do not occur.

\subsubsection{$K$-corrections to the Radial Velocity Curves}

In X-ray binaries, the secondary star is not
uniformly bright over its surface.  Even for 
an unirradiated star, gravity darkening causes 
the star to look dimmer near the inner Lagrangian point.  
If the star is irradiated by the X-ray source, then the hemisphere 
facing the source appears brighter.  In either 
case, the ``center of light'' as seen in sky 
coordinates may not coincide with the star's 
center of mass.  At a given time, the observed 
profile of a spectral line is the flux-weighted 
and Doppler-shifted sum of the individual profiles 
distributed over the visible portion of the star's surface.  
The brightness variations over the surface cause asymmetries in 
the spectral line profiles, which in turn cause the measured radial 
velocity to differ from the true radial velocity (Wilson \& Sofia 1976).  
For the simple case of a circular orbit, e.g., this causes the 
velocity curve to deviate from a sinusoid, and the 
measured $K$-velocity to differ from the true $K$-velocity.  
In practice, to obtain the true $K$-velocity one computes the so-called 
``$K$-correction'' and adds it to the observed $K$-velocity.

\begin{figure}[t!]
\includegraphics[scale=0.3,angle=-90]{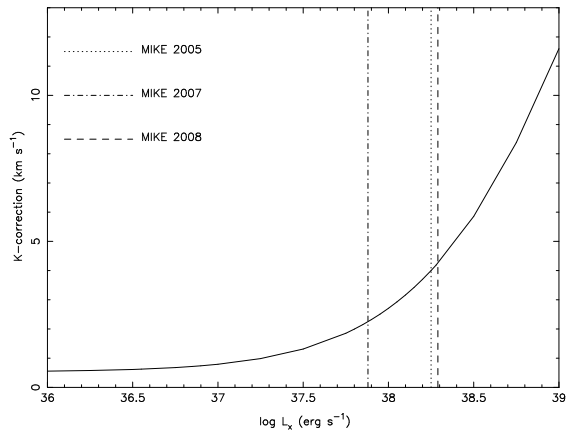}
\caption{The $K$-correction as a function of the X-ray
luminosity.  The vertical lines give the X-ray luminosity
during each of the three MIKE runs.}
\label{plotcor}
\end{figure}

\begin{figure*}[t!]
\includegraphics[scale=0.62,angle=-90]{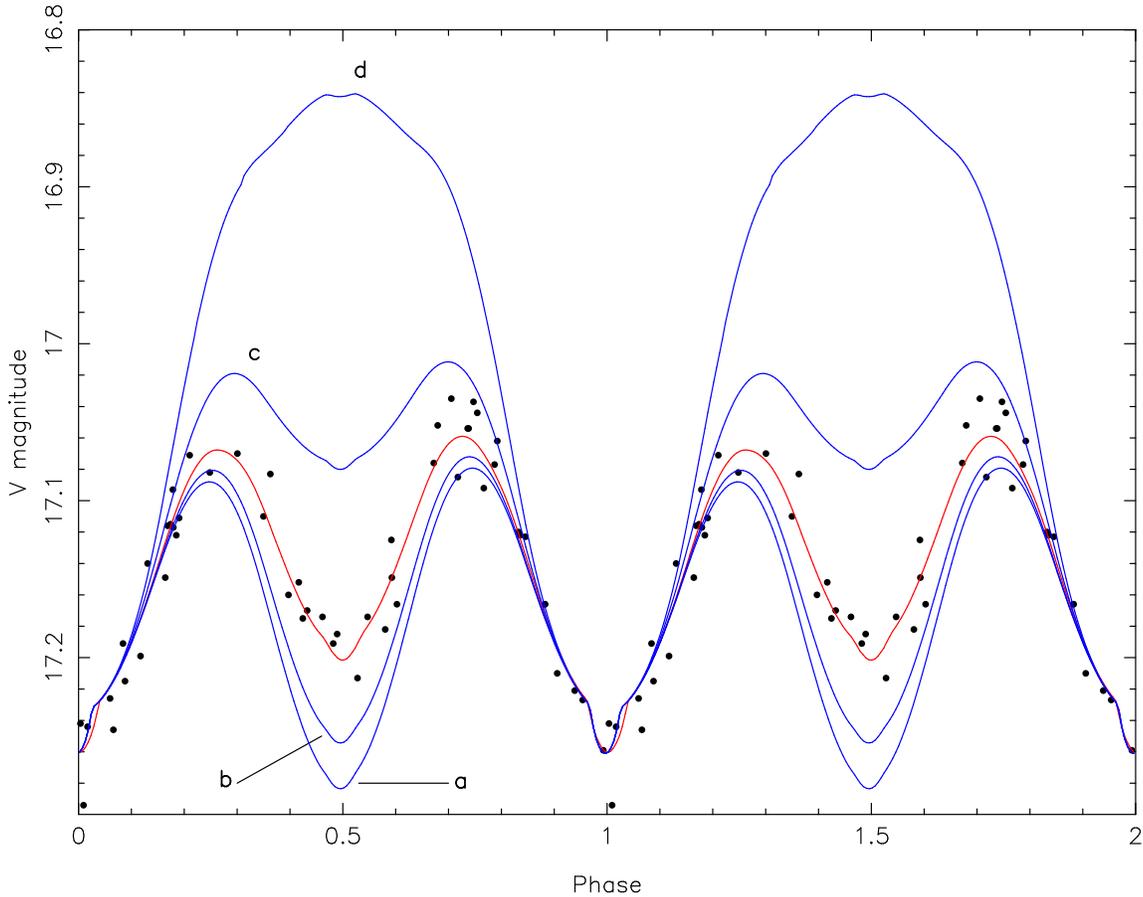}
\caption{The $V$-band light curve from the W95 data set
is shown along with the best-fitting model (red line)
with $\llog=38$.  
This model includes a grazing eclipse of the accretion 
disk at phase 0; we do not claim that such a 
feature is definitively present in the data.
Four other models with different X-ray heating
but otherwise identical parameters are shown in blue.
These models are (a) no X-ray heating;
(b) $\llog=37.5$
and a disk-like X-ray emitting emitting
region; (c)   
$\llog=38.5$
and a disk-like X-ray emitting emitting region;
and (d) $\llog=38.0$
and a point-like X-ray emitting emitting
region.}
\label{heatfig1}
\end{figure*}

The $K$-correction can be computed in two distinctly different ways.
If the X-ray spectrum is ``soft'', the X-rays are absorbed
near the stellar surface and the absorption lines are
``filled in'' and are hence considerably weaker than normal
(Phillips et al.\ 1999;  Soria et al.\ 2001).
Since in this case the absorption lines one
actually observes come mostly from the unheated
part of the star, the true $K$-velocity
is {\em smaller} than the measured one, and
the $K$-correction is {\em negative}.
If, on the other hand, the X-ray
spectrum is ``hard'', the X-rays are either absorbed below the photosphere
or are reflected, which strengthens the absorption lines.  In this case,
the dominant contribution to the observed
absorption lines comes from the heated
part of the star and, consequently, the true $K$-velocity is {\em larger}
than the measured one, and the $K$-correction is 
{\em positive}.

In two ways our data suggest that the $K$-correction is positive and corresponds 
to the ``hard'' case outlined above.
First, we observe a change of temperature over the 
orbit, with the hottest temperatures observed when the inner Lagrangian
point is pointed at the observer.  If absorption lines were largely absent
from that part of the star, then we should not expect to measure
such a temperature change.  Second, the three MIKE data sets show
an anti-correlation between the X-ray luminosity and the $K$-velocity
(Table 1).  If one ranks the three MIKE runs in terms of the
X-ray luminosity from the lowest to the highest, the ordering is
2007, 2005, and 2008.  If one arranges the $K$-velocities from
the largest to the smallest, the ordering is again 2007, 2005, and 2008.
The $K$-velocity from the UVES data does not fit this trend.  However,
the X-ray luminosity changed significantly between the third and
fourth UVES observation.  
Also, with only five observations the phase coverage is poor.

ELC computes corrections to the model radial velocity curve
following the prescription given in Wilson \& Sofia (1976).  This corresponds
to the ``hard'' case discussed above, which results in positive $K$-corrections.
To compute the $K$-corrections, we used the W95 light curve solution as the
base model (note that this model includes an accretion disk).
The choice of the W95 light curves was somewhat arbitrary, but the 
results are insensitive to the actual data set used.  
Light and velocity curves were computed using a wide range of
X-ray heating values, and sine curves were fitted to the model curves to
measure the model $K$-velocity.  The $K$-correction is defined as
the input $K$-velocity minus the fitted $K$-velocity.  Figure
\ref{plotcor} shows the results.  Note that there is still a
small $K$-correction (0.56 km s$^{-1}$) when the X-ray luminosity is 
small compared to the bolometric luminosity of the star---this is due
to the  tidal
distortions of the star.  The $K$-correction is about 5 km s$^{-1}$
when $\llog=38.5$, which is at the upper end of the range of the
observed X-ray luminosity of the source.

In computing the $K$-corrections for the four 
radial-velocity data sets, we estimated the 
X-ray luminosity using the mean {\em RXTE} ASM 
count rates and the standard deviation of the 
individual ASM measurements (see Table 1 for count 
rates and luminosities).  The count rate was nearly 
constant for the 2007 and 2008 MIKE runs, somewhat variable 
for the 2005 MIKE run and highly variable for the 
UVES run.  For a given data set, the standard deviation 
in the count rate was added in quadrature with the 
mean uncertainty on an individual measurement to produce 
the adopted uncertainty on the ASM count rate.  
Finally, the mean ASM count rate was converted to an X-ray 
luminosity and the $K$-correction was computed using the 
curve shown in Figure \ref{plotcor}.  The $K$-corrections and the 
values of the final corrected $K$-velocities are given in Table 1.

The typical X-ray luminosity of LMC X-3 ($L_{\rm x}\gtrsim 5\times 10^{37}$
erg s$^{-1}$) exceeds the bolometric luminosity of the
star ($L_{\rm bol}\sim 4\times 10^{36}$ erg s$^{-1}$) by more
than an order of magnitude, so one might expect the effects
of X-ray heating to be large.
Nevertheless, 
the $K$-corrections are modest, with the largest one being $4.47\pm 0.86$
km s$^{-1}$ for the UVES data.  
In addition, the light curves are
distinctly double-waved, which also points to a modest
amount of X-ray heating.  
In LMC X-3, there are two main reasons why the X-ray heating is
weak.  First, the geometry of the X-ray emitting area has
a disk-like structure rather than a point-like structure.  With
a disk-like structure, the X-ray source appears to be 
somewhat foreshortened when viewed from the secondary star.  Second,
the outer parts of the accretion disk can block the X-rays
from hitting parts of the secondary star that are near its equator.
The need for a disk-like geometry for the X-ray emitting area
is shown in Figure \ref{heatfig1}, which  displays the
$V$-band light curve from W95 and the best-fitting
model with $\llog=38$.  For comparison we
show other models with
different X-ray luminosities but identical parameters otherwise.
The X-ray luminosity cannot be much larger than $\sim 10^{38.1}$
erg s$^{-1}$
since the depth of the minimum near phase 0.5 becomes
too shallow. 
We also show the model with $\llog=38$
and a point-like geometry.
This model can be clearly ruled out by the data since it has a single
maximum and a single minimum per orbital cycle.

For our final adopted value of the $K$-velocity, we begin
by taking the
average of the four individual measurements given in Table 1.
We find $K_2=241.1\pm 2.8$ km s$^{-1}$, where the quoted
uncertainty is the error of the mean.  To account for 
possible systematic errors caused by X-ray heating, we add
the standard deviation of the four measurements
(5.5 km s$^{-1}$) in quadrature to the formal error in
the mean to produce our adopted uncertainty of 6.2 km s$^{-1}$,
thereby giving $K_2=241.1\pm 6.2$ km s$^{-1}$.

\subsubsection{Adopted Results}

To arrive at our final adopted result, 
we first find the weighted
average of the inclination $i$ from the 13 data sets for both the models
with X-ray heating and without.  {The adopted weights for the
individual inclination measurements are taken to be $W=1/\sigma^2$.  
Furthermore, }
when computing the weighted averages,
we impose a ``floor'' on the inclination uncertainty of $0.5^{\circ}$ to
avoid giving undue weight to those few cases with extremely small
uncertainties.  As the uncertainty on the adopted inclination, we use
the dispersion-adjusted error in the mean.  
Then, given $i$ and its
uncertainty, we compute the mass of the black hole, the mass and radius
of the secondary star, and other parameters assuming a $K$-velocity of
$K_2=241.1\pm 6.2$ km s$^{-1}$, an orbital period of $P=1.7048089$
days, and a rotational velocity of $V_{\rm rot}\sin i=118.5 \pm 6.6$ km
s$^{-1}$.  
The computation of the mass ratio $Q$ via $V_{\rm rot}\sin i/K_2$ 
involves accurate numerical integrations in Roche geometry.
The final
quantities are summarized in Table \ref{resultstab} separately for the
two models which respectively omit or incorporate treatment of X-ray
heating.  
Given that the models which include X-ray heating are more complete, we 
adopt the values derived for this case.
In particular, we find $M=6.98\pm 0.56\,M_{\odot}$.

\begin{figure}[t!]
\epsscale{1.}
\plotone{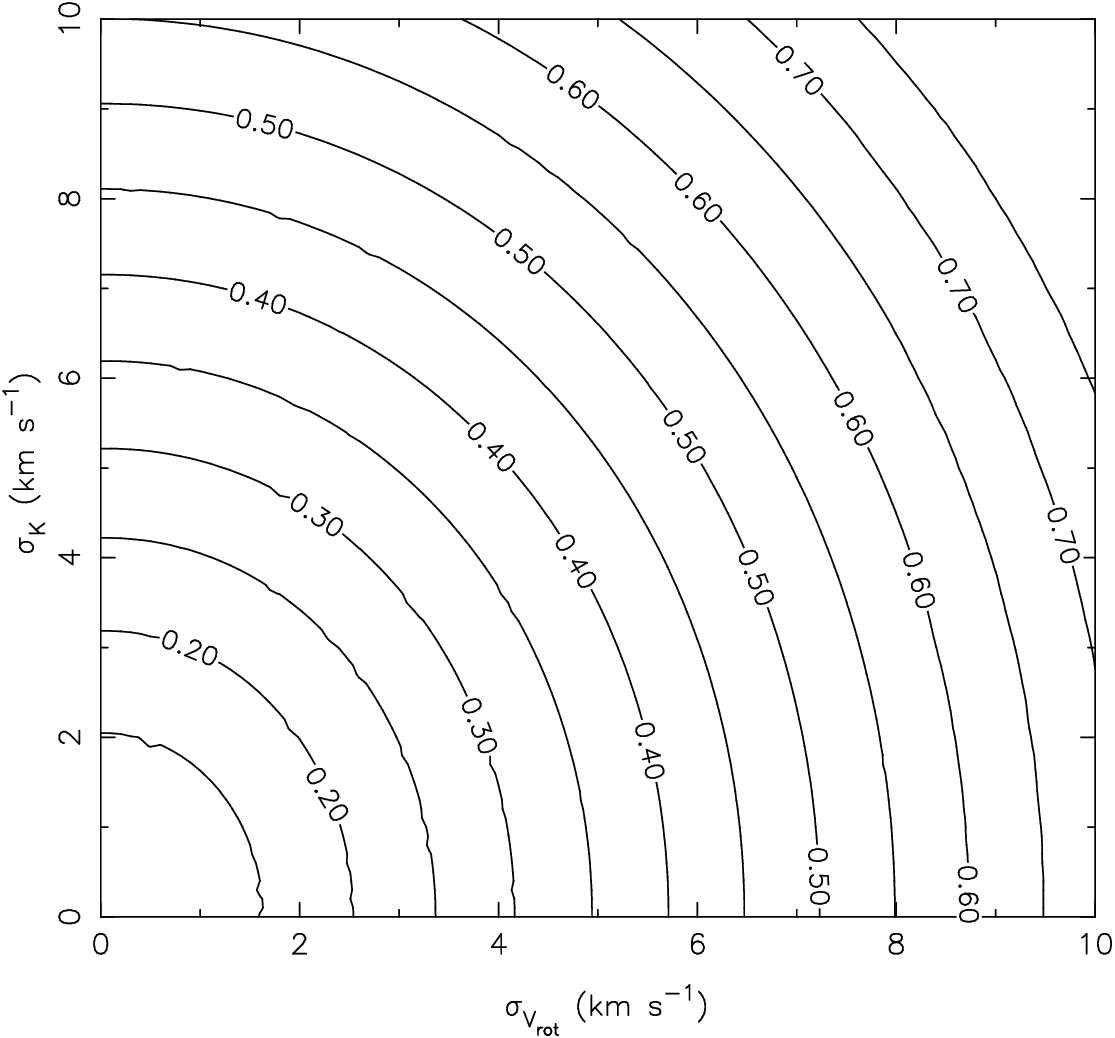}
\caption{The uncertainty on the black hole mass
(in $M_{\odot}$)
as a function of the uncertainty on $V_{\rm rot}\sin i$
($x$-axis) and the uncertainty on the $K$-velocity
($y$-axis).  The contour levels are spaced 
every $0.05\,M_{\odot}$, but for clarity only the levels
0.10, 0.20, 0.30, etc. are labeled. }
\label{errorcontour}
\end{figure}

\setcounter{table}{9}
\begin{deluxetable}{lrr}
\tablecaption{Final Parameters of LMC X-3\label{resultstab}}
\tablewidth{0pt}
\tablehead{
\colhead{parameter} &
\colhead{(No heating)} &
\colhead{(X-ray heating)}}
\startdata
$i$ (deg) &  $69.84 \pm 0.37$  &  $69.24\pm 0.727$\tablenotemark{a} \cr
$Q$  &  $1.93\pm 0.20$  & $1.93\pm 0.20$\tablenotemark{a} \cr
$a$ ($R_{\odot}$) & $13.08 \pm 0.44$ & $13.13\pm 0.45$\tablenotemark{a} \cr
$M_2$ ($M_{\odot}$)  &  $ 3.58\pm 0.56$ & $3.63\pm 0.57$\tablenotemark{a} \cr 
$R_2$ ($R_{\odot}$)  &  $ 4.23 \pm  0.24$  & $4.25\pm 0.24$\tablenotemark{a}\cr
$\log g_2$ (cgs)  &  $ 3.739 \pm  0.020$  & $3.740\pm 0.020$\tablenotemark{a}  \cr
$M$ ($M_{\odot}$)  &  $ 6.90\pm 0.55 $  &  $6.98\pm 0.56$\tablenotemark{a} \cr
\enddata
\tablenotetext{a}{Adopted value.}
\end{deluxetable}

Figure \ref{errorcontour} shows how the uncertainty on
the black hole mass depends on the uncertainties of
the rotational velocity and the $K$-velocity.  The
uncertainties in the inclination and the period are
included in the computations.
When $\sigma_{V_{\rm rot}}=6.6$ km s$^{-1}$,
the uncertainty in the black hole mass changes modestly
for $\sigma_K\lesssim 6$ km s$^{-1}$, since
the contour lines are nearly perpendicular to the
horizontal axis of the contour plot.  Thus most of
the  improvement
in the accuracy of the black hole mass determination must
come from an improvement in the determination of
the rotational velocity.

\subsection{Consistency Check Using the Distance to the LMC}

Previously, we have made use of a well-determined distance to the source
to put constraints on the dynamical models used to find the black hole
mass (and other system parameters) in the high mass X-ray binaries M33
X-7 (Orosz et al.\ 2007), LMC X-1 (Orosz et al.\ 2009), and Cyg X-1
(Orosz et al.\ 2011).  
Unlike LMC X-3's secondary, the companion stars in these systems do not 
fill their Roche lobes, and therefore the 
ellipsoidal light curves by themselves are much less constraining.
Fortunately, for these systems the radius of the companion star can be
found from the distance, the apparent magnitude, the extinction, the
effective temperature of the star, and the bolometric correction
determined from model atmospheres.  
The independently derived stellar radius 
in turn serves to strongly constrain the 
available parameter space for the dynamical model.

\begin{figure}[t!]
\epsscale{1.}
\plotone{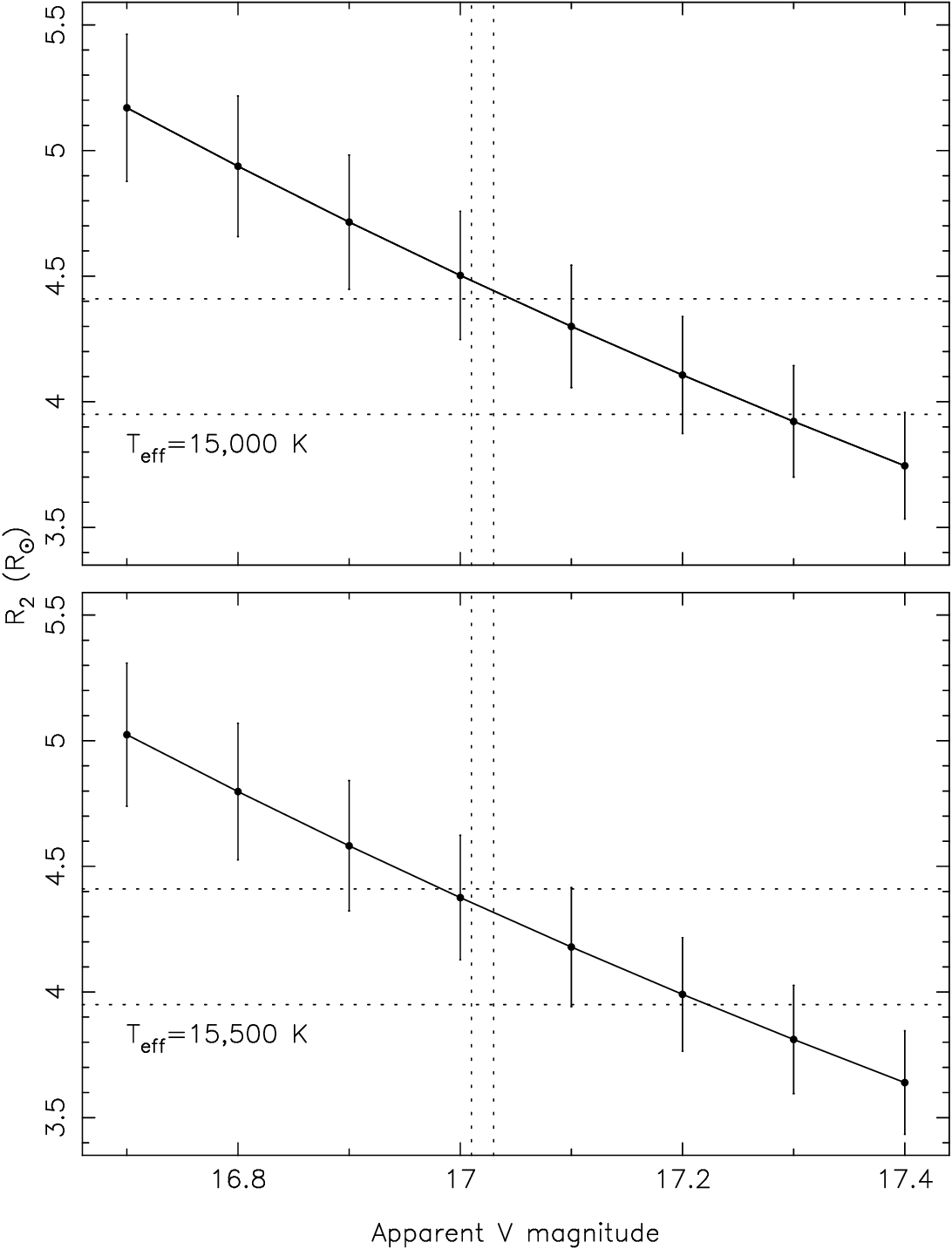}
\caption{The computed radius of the secondary
star as a function of the apparent $V$ magnitude
assuming an effective temperature of
$T_{\rm eff}=15,000\pm 100$~K (top) and
$T_{\rm eff}=15,500\pm 100$~K (bottom).
We assume a distance modulus of $18.41\pm 0.10$, an
extinction of $A_V=0.223\pm 0.030$, and a disk fraction of
$0.15\pm 0.05$.  The horizontal dotted lines denote the $1\sigma$ range
in the radius found from the dynamical model, and the vertical
dotted lines denote the $1\sigma$ range of apparent $V$ magnitude of
the system when the accretion disk light is minimal.
}
\label{radvsV}
\end{figure}

By virtue of its membership in the LMC (e.g.\ see Cowley et al.\ 1983),
the distance to LMC X-3 is likewise well determined, and we adopt a
distance modulus of $18.41\pm 0.10$ mag as documented in the Supplementary
Information of Orosz et al.\ (2007).  However, given the erratic optical
variability that is added to the underlying ellipsoidal modulation and
the presence of light from the accretion disk, it is hard to define a
baseline apparent magnitude for LMC X-3.  Instead we take a different
approach.  For a fixed apparent $V$ magnitude, we can compute what the
radius of a spherical star would have to be at the distance of the LMC
given a temperature, extinction, a bolometric correction, and a
correction for light from the accretion disk.  For this exercise we used
a $V$-band extinction of $A_V=0.223\pm 0.030$ based on the column
density of $N_H=(3.8\pm 0.8)\times 10^{20}$ cm$^{-2}$
(Page et al.\ 2003) and the
conversion from $N_H$ to $A_V$ given by Predehl \& Schmitt (1995).  The
bolometric corrections were interpolated from the BSTAR 2006 grid given
by Lanz \& Hubeny (2007).  We assume the star contributes $85\pm 5\%$ of
the $V$-band light.  Using these assumptions, we computed the radius and
its uncertainty for a range of apparent $V$ magnitudes for an effective
temperature of $T_{\rm eff}=15,000\pm 100$ K and for $T_{\rm
eff}=15,500\pm 100$ K.  The results are shown in Figure \ref{radvsV}.

For a fixed $V$ magnitude, the formal uncertainty in the computed radius
is between about 0.2 and $0.3\,R_{\odot}$, compared to the uncertainty
of $0.24\,R_{\odot}$ from the dynamical model.  The $1\sigma$ regions of
the  model radius (shown as the horizontal dotted lines in
Figure \ref{radvsV}) and the radius from the distance overlap for
$17.05\lesssim V\lesssim 17.30$ when $T_{\rm eff}=15,000$ K and
$16.95\lesssim V \lesssim 17.20$ when $T_{\rm eff}=15,500$ K, with the
best matches occurring for $V\approx 17.15$ at $T_{\rm eff}=15,000$ K and
$V\approx 17.10$ at $T_{\rm eff}=15,500$ K.  

This range of V magnitude for the star is in good agreement with the
minimum V magnitude that we infer for LMC X-3 in two ways using the
well-calibrated SMARTS data: (1) For the ``X-quartered'' data for which
X-ray heating is minimal, the apparent $V$ magnitude is 17.1 at the
minimum at phase 0.5 and 16.95 at the maximum at phase 0.25; a spherical
star with a radius equal to the effective radius of the companion star
would have $V=17.02$.  (2) Focusing on the anomalous low state of 2008
December 11 -- 2009 June 17 reported by Smale \& Boyd (2012), and
allowing for the two-week lag in X-ray intensity, the mean and standard
deviation for 24 consecutive, near-nightly SMARTS observations is $V =
17.05\pm 0.08$.  Adopting as our minimum $V=17.05$ and allowing for a 1\%
uncertainty in the zero-point, the vertical dotted lines in Figure
\ref{radvsV} show the $1\sigma$ range of the apparent $V$ magnitude of
the system.  The vertical and horizontal dotted lines form a box in the
$V$-$R_2$ plane, and the curve defining the computed radius as a
function of the distance passes through this box when $T_{\rm
eff}=15,500$ K and passes very close when $T_{\rm eff}=15,000$ K.

There are claims that LMC X-3 is at times fainter by a few tenths of a
magnitude than our adopted minimum of $V = 17.05$~mag (Van der Klis et
al. 1983; Soria et al. 2001; Brocksopp et al. 2001).  However, the
$V$-band zero point is much less secure for the data considered in this
earlier work than for our SMARTS data, where it was determined via
standard star observations on 44 photometric nights. 
Furthermore, it seems unlikely that LMC X-3 was actually fainter in
these earlier observations given that our determination was made during
an extreme low state (Smale and Boyd 2012).

Thus, to summarize, the radius of the star computed using the distance,
temperature, apparent magnitude, and extinction is fully consistent with
the radius found from the dynamical model when $T_{\rm eff}=15,500$ K
and is well within the $1\sigma$ uncertainty when $T_{\rm eff}=15,000$
K.  
The strong consistency in the radius derived from the two different
methods is reassuring and gives us extra confidence in our
dynamical model.

\section{Discussion and Conclusions}

Our mass for the black hole of $6.98\pm 0.56\,M_{\odot}$ is considerably
more precise than earlier estimates for three principal reasons: First,
we have far more radial velocity data of high quality for the secondary
star, which yield a much improved determination of its $K$-velocity.
Second, we have obtained the first accurate measurement of the 
projected rotational velocity of the secondary, which is an important constraint on the dynamical model.
Third, we have analyzed a much larger body of
ellipsoidal light curve data than any previous study (e.g., Kuiper et
al. 1988).

In their pioneering work on LMC X-3, Cowley et al. (1983) concluded that
the most plausible mass of the black hole was $\sim 9\,M_{\odot}$ with a
lower limit of $M>7\,M_{\odot}$.  To reach this conclusion, Cowley et
al.\ had to assume a mass for the secondary star.  Based on their
spectral classification of the star as B3V, they deduced its mass to be
$4-8\,M_{\odot}$, greater than our value of $M_2=3.63\pm
0.57\,M_{\odot}$.  Estimating the mass based on the spectral type is
problematic for a star that has lost an indeterminate amount of mass.
On a related note, our measurement of the surface gravity of the star,
$\log g=3.740 \pm 0.020$, shows that it matches more closely the gravity
of a giant (class III) than the gravity of a main sequence star (class
V).  For example, the nominal masses and radii for a B3V/B5V star 
given by Ostlie \& Carroll (1996) correspond to surface gravities of
$\log g=4.07/3.98$, while their surface gravity for a B5III star is
$\log g=3.68$.

Soria et al.\ (2001) and Val-Baker et al.\ (2007) provided mass
estimates after making corrections to the $K$-velocity of the secondary
star to account for X-ray heating.  Based on the system colors as
measured by the Optical Monitor on {\em XMM-Newton}, Soria et al.\
concluded that the secondary star is a B5 subgiant that fills its Roche
lobe.  Using an evolutionary model they determined a mass of
$M_2=4.7\,M_{\odot}$ for the secondary star.  Next, Soria et al.\
applied a $K$-correction of $\Delta K=-30\pm 5$ km s$^{-1}$ to the
$K$-velocity from Cowley et al.\ (1983) to arrive at the ``true''
$K$-velocity of $205\pm 12$ km s$^{-1}$.  
The sign of their $K$-correction is negative because 
Soria et al. used the procedure that assumes 
heating by ``soft'' X-ray photons (Section 5.2.2).
Finally, Soria et al.\ used their
secondary star mass and corrected $K$-velocity to arrive at their
lower limit of $M > 5.8\pm 0.6\,M_{\odot}$.

Val-Baker et al.\ (2007) measured a $K$-velocity of $K_2=242.4\pm 4.3$
km s$^{-1}$ from the five UVES spectra.  
Their $K$-correction of 14.3 km s$^{-1}$ is 
positive because they used the procedure that 
assumes irradiation by ``hard'' photons; their 
adopted $K$-velocity is therefore $K_2=256.7\pm 4.9$ km s$^{-1}$ .
By comparing their spectra
to standard star spectra, Val-Baker et al.\ determined a spectral type
of B3V at phase 0.22 when X-ray heating is significant, and B5V at phase
0.86, when X-ray heating is slight.  Converting these spectral types
into temperatures using Kurucz models, they arrived at effective
temperatures of 15,400 K for the unheated face and 18,700 K for the
heated face.  This temperature change was used to then find
their $K$-correction.  Finally, after
adopting a nominal mass of $5.9\,M_{\odot}$ for a B5V star
(e.g.\ Ostlie \& Carroll 1996), Val-Baker found a mass range for the
black hole of $9.5\le M \le 13.6\,M_{\odot}$.

Soria et al.\ (2001) and Val-Baker et al.\ (2007) not only disagree by a
factor of two on the magnitude of the $K$-correction,
they even disagree on its sign.  As noted earlier,
the disagreement on the sign of the
effect results from Soria et al.\  assuming that the X-ray spectrum
illuminating the star is soft, while Val-Baker et al.\ assumed it is hard.
Setting aside the difference in the signs for 
the moment, we believe that the magnitudes of 
these corrections are too large [where $|\Delta K|=30$ 
km s$^{-1}$ for Soria et al.\ (2001) and $|\Delta K|=14.3$ 
km s$^{-1}$ for Val-Baker et al.\ (2007)].  
In contrast, consideration of our much more 
extensive data set shows that X-ray heating 
has a modest effect on the $K$-velocity (Section 4.1), 
which is probably because the star is shielded 
by the outer part of the accretion disk and 
because the X-ray emitting region has a disk-like geometry.
Val-Baker et al.\ inferred a temperature
difference of 3300 K between the heated and unheated faces of the star.
However, as shown in Figure
\ref{temperaturetrend}, we measure a temperature range of less   
than half their value.  The result of Val-Baker et al.\ is subject to the
complication that the X-ray luminosity of the system (as measured by the
{\em RXTE} ASM) varied by a factor of two over the course of their UVES
observations (see Table \ref{BAT}).  
Because the $K$-correction is strongly supralinear with temperature,
our correction for
1500 K is much smaller than that of Val-Baker et al. for 3300K.

The black hole mass of $6.98\pm 0.56\,M_{\odot}$ is entirely consistent
with the masses of black hole in transient systems with low-mass
secondaries ($7.8 \pm 1.2~M_{\odot}$; 
\"{O}zel et al. 2010; Farr et
al. 2011). In contrast, the black hole masses in the
persistently-bright, wind-fed systems with high mass secondaries (e.g.\
M33 X-7, LMC X-1, and Cyg X-1) are considerably larger (Orosz et al.\
2007, 2009, 2011).  Considering the mass of the black hole and the
following three other properties of LMC X-3, we classify the system as a
member of the transient black hole binaries that are fed by Roche-lobe
overflow, rather than as a member of the persistent systems whose black
holes are wind-fed by massive O-type or Wolf-Rayet secondaries
(McClintock et al. 2013): (1) Mass transfer via Roche-lobe overflow is
expected given the mass ratio reported herein ($M/M_2 \approx 2$).
(2) Although normally X-ray bright, LMC X-3 is highly variable and on
occasion enters a prolonged low-intensity state that is dominated by a
hard power-law component (Wilms et al. 2001; Smale \& Boyd 2012).  (3)
Likewise, the low spin of the black hole (Davis et al. 2006; Steiner et
al. 2014) distinguishes LMC X-3 from the persistent black holes and is
consistent with that of other transient black holes (McClintock et
al. 2013).

One obvious difference between LMC X-3 and a typical transient system
like A0620-00 is that LMC X-3 has never been reported to be in a true
X-ray quiescent state with $L_{\rm x}\lesssim 10^{33}$ ergs s$^{-1}$.  The
mass transfer rate in LMC X-3 is high compared to a typical transient
because the relatively massive secondary is presumably evolving on a
nuclear timescale.  The higher mass transfer rate results in much
shorter recurrence timescales, according to the disk instability model
(Cannizzo, Chen, \& Livio 1995; Lasota 2001).  In the context of
this model, perhaps the recurrence time is so short that LMC X-3 is
perpetually in an outburst state.

Finally, although a detailed evolutionary model is beyond the scope of
this paper, we briefly speculate on the future of LMC X-3.  Since the
present-day mass of the secondary star is less than that of the black
hole, the mass transfer is expected to be both thermally and dynamically
stable (Tauris \& van den Heuvel 2006).  Mass transfer from the lower
mass object to the higher mass object results in an expansion of the
orbit (e.g., the period and the separation both increase).  The
secondary star's radius would also increase, as it continues to fill its
Roche lobe.  The mass that is transferred through the accretion disk
will tend to increase the spin of the black hole.  By the time core
H-burning ceases, the star may have transferred a considerable amount of
its current mass to the black hole.  Thus, it appears that LMC X-3, will
end up with a relatively massive black hole, a long orbital period,
appreciable spin, and a low mass companion.  This future version of LMC
X-3 will likely bridge the gap between most transient systems
and the extreme system GRS 1915+105 (McClintock et al.\ 2006; Steeghs et
al.\ 2013).

\acknowledgments

JFS was supported by NASA Hubble Fellowship grant HST-HF-51315.01.  The
work of JEM was supported in part by NASA grant NNX11AD08G.  DS
acknowledges a STFC Advanced Fellowship as well as support through the
NASA Guest Observer Program.  MMB and CDB acknowledge support from the
National Science Foundation via the grants AST 0407063 and AST 070707.

{ 
\appendix

\section{Conversion of {\em RXTE} ASM fluxes to X-ray Luminosities}

To convert a measurement of the 2-12 keV X-ray intensity from the {\em RXTE} 
All Sky Monitor 
(ASM) into an estimate of the bolometric X-ray luminosity, 
the following procedure was used.
First, the RXTE ASM count rate was converted into a Crab-unit
equivalent intensity, and subsequently into a flux using the 2-12 keV
brightness of the Crab, which is a commonly employed flux conversion
method. 
Each flux measurement was then converted to an X-ray luminosity using
$L_x ({\rm erg}\,\, {\rm s}^{-1})=
({\rm ASM\, rate})\times
(8.85\times10^{37})$, assuming isotropic emission and a distance of
$d=48.1$ kpc.
We do not use a particular model in making this conversion and
in deriving the bolometric correction; rather, we take the following
approach:

1) We consider fits to typical low-and-high luminosity RXTE PCA
spectra of LMC X-3 consisting of a disk and power-law ({\tt simpl}
$\times$ {\tt kerrbb};
see Steiner et al.\ 2010).  We find that the model-extrapolated
bolometric (10 eV -- 100 keV) flux is typically two to three 
times larger
than the 2-12 keV flux, with a bolometric conversion 
factor of 2.5 being a
good and representative value.

2) We have estimated the error induced from having computed fluxes
based upon the spectral shape of the Crab.  Specifically, 
we checked limiting cases in which LMC X-3’s spectrum is
assumed to be i) a pure blackbody and ii) a pure power-law. In the
first instance, for characteristic temperatures between 0.1-1.0 keV,
we find
that our conversion {\em overestimates} the X-ray flux by 
10 to 20\%.  
In the
second instance, we examine spectral indexes ranging from 2.0-2.7 (see
Steiner et al.\ 2010), and find that the flux error 
is  $\lesssim 10\%$.
These errors are commensurate with our typical count rate
uncertainties, which are $\sim 20\%$, 
indicating that although a bias may be present,
it would manifest at the $\sim 1\sigma$ level.  
Furthermore, we note that
since we do not have the instantaneous X-ray spectral information on
the system which would enable a precise flux estimate, any different
model-based approach would be subject to comparable systematic
uncertainties, but it would also be more convoluted than our present
approach, which has the virtue of being simple and easily reproduced.

}

\twocolumn



\begin{deluxetable}{ccccrrrr}
\setcounter{table}{1}
\tablecaption{Stellar parameters from spectra of  LMC X-3}
\tabletypesize{\scriptsize}
\tablewidth{0pt}
\tablehead{
\colhead{UT date} &
\colhead{HJD}  &
\colhead{phase} &
\colhead{ASM\tablenotemark{a}} &
\colhead{$T_{\rm eff}$} &
\colhead{$\log g$} &
\colhead{$V_{\rm rot}\sin i$} &
\colhead{$k$}    \\
\colhead{(YYYY-MM-DD)} &
\colhead{(+2,450,000)} & 
\colhead{} &
\colhead{ct s$^{-1}$} &
\colhead{(K)} &
\colhead{(cgs)} & 
\colhead{(km s$^{-1}$)}&
\colhead{}}
\startdata
\multicolumn{8}{c}{UVES Spectra}\\
\hline
2004-12-19 &3358.56490  &0.860 & $1.82\pm0.42$ & $15,400\pm 47$  & $3.695\pm 0.030$ & $97.8\pm 3.3$ & $0.797\pm 0.001$\\
2005-01-04 &3374.72113  &0.336 & $1.89\pm0.15$ & $16,750\pm 120$ & $3.792\pm 0.032$ & $119.9\pm 4.4$ & $0.769\pm 0.003$\\
2005-01-08 &3378.63666  &0.633 & $1.87\pm0.30$ & $16,541\pm 133$ & $3.868\pm 0.008$ & $111.0\pm 4.1$ & $0.755\pm 0.003$\\
2005-03-19 &3448.52084  &0.626 & $3.15\pm0.18$ & $16,682\pm 127$ & $3.812\pm 0.036$ & $120.9\pm 4.0$& $0.676\pm 0.002$ \\
2005-03-20 &3449.53491  &0.220 & $3.14\pm0.54$ & $16,326\pm 96$ & $3.784\pm 0.031$ & $143.1\pm 2.9$ & $0.747\pm 0.003$\\
\nodata & \nodata& \nodata  &\nodata & $16,435\pm  60$\tablenotemark{b} 
                         & $3.801\pm 0.010$\tablenotemark{b}  
                   &  $121.4\pm 1.4$\tablenotemark{b} 
             & $0.732\pm 0.001$\tablenotemark{b}  \\
\hline
\multicolumn{8}{c}{MIKE Spectra} \\
\hline
2005-01-20    &        3390.74244   & 0.734     &            $1.93\pm0.53$ &$ 16,211\pm    80$  &  $3.541 \pm 0.009 $  &  $131.7 \pm  4.3$ & $ 0.720 \pm 0.004$ \\
2005-01-20    &        3390.77549   & 0.754     &            $1.94\pm0.54$ &$ 15,882\pm   105$  &  $3.685 \pm 0.030 $  &  $118.3 \pm  7.0$ & $ 0.724 \pm 0.004$ \\
2005-01-21    &        3391.68953   & 0.290     &            $1.96\pm0.65$ &$ 15,992\pm   184$  &  $2.885 \pm 0.018 $  &  $158.7 \pm  6.1$ & $ 0.702 \pm 0.009$ \\
2005-01-22    &        3392.77873   & 0.929     &            $1.99\pm0.53$ &$ 15,220\pm   180$  &  $3.481 \pm 0.050 $  &  $100.8 \pm 10.7$ & $ 0.609 \pm 0.006$ \\
2005-01-24    &        3394.69509   & 0.053     &            $2.03\pm0.37$ &$ 15,722\pm   327$  &  $2.985 \pm 0.030 $  &  $190.4 \pm  6.9$ & $ 0.721 \pm 0.015$ \\
2005-01-25    &        3395.73754   & 0.664     &            $2.08\pm0.55$ &$ 16,278\pm   321$  &  $3.427 \pm 0.014 $  &  $141.9 \pm  7.6$ & $ 0.711 \pm 0.008$ \\
\nodata & \nodata& \nodata  &\nodata & $15,781\pm  61$\tablenotemark{b} 
                         & $3.319\pm 0.004$\tablenotemark{b}  
                   &  $132.6\pm 2.9$\tablenotemark{b} 
             & $0.664\pm 0.002$\tablenotemark{b}  \\
\hline
2007-12-20    &        4454.58242   & 0.757     &            $0.85\pm0.31$ &$ 15,762\pm   104$  &  $3.692 \pm 0.024 $  &  $121.4 \pm  4.4$ & $ 0.873 \pm 0.003$ \\
2007-12-20    &        4454.60559   & 0.771     &            $0.85\pm0.31$ &$ 15,964\pm    43$  &  $3.862 \pm 0.020 $  &  $121.3 \pm  4.7$ & $ 0.875 \pm 0.001$ \\
2007-12-20    &        4454.62818   & 0.784     &            $0.85\pm0.31$ &$ 15,440\pm    60$  &  $3.650 \pm 0.014 $  &  $123.3 \pm  5.1$ & $ 0.853 \pm 0.002$ \\
2007-12-20    &        4454.65045   & 0.797     &            $0.85\pm0.31$ &$ 15,635\pm    95$  &  $3.775 \pm 0.027 $  &  $117.3 \pm  5.1$ & $ 0.882 \pm 0.002$ \\
2007-12-20    &        4454.68433   & 0.817     &            $0.85\pm0.31$ &$ 15,497\pm   154$  &  $3.777 \pm 0.028 $  &  $118.2 \pm  4.2$ & $ 0.863 \pm 0.005$ \\
2007-12-20    &        4454.70568   & 0.829     &            $0.85\pm0.31$ &$ 15,418\pm   109$  &  $3.751 \pm 0.016 $  &  $116.7 \pm  3.2$ & $ 0.860 \pm 0.003$ \\
2007-12-20    &        4454.72792   & 0.843     &            $0.85\pm0.31$ &$ 15,207\pm    71$  &  $3.692 \pm 0.015 $  &  $108.7 \pm  3.3$ & $ 0.867 \pm 0.002$ \\
2007-12-20    &        4454.75014   & 0.856     &            $0.85\pm0.31$ &$ 15,398\pm   115$  &  $3.710 \pm 0.023 $  &  $113.7 \pm  4.9$ & $ 0.884 \pm 0.003$ \\
2007-12-20    &        4454.77178   & 0.868     &            $0.85\pm0.31$ &$ 15,090\pm   115$  &  $3.580 \pm 0.023 $  &  $123.6 \pm  3.8$ & $ 0.879 \pm 0.004$ \\
2007-12-20    &        4454.79457   & 0.882     &            $0.85\pm0.31$ &$ 15,381\pm    76$  &  $3.655 \pm 0.021 $  &  $111.0 \pm  3.6$ & $ 0.873 \pm 0.003$ \\
2007-12-20    &        4454.81744   & 0.895     &            $0.85\pm0.31$ &$ 15,207\pm    71$  &  $3.721 \pm 0.021 $  &  $107.7 \pm  3.8$ & $ 0.872 \pm 0.002$ \\
2007-12-20    &        4454.83678   & 0.906     &            $0.85\pm0.31$ &$ 15,018\pm   151$  &  $3.584 \pm 0.024 $  &  $117.2 \pm  6.4$ & $ 0.897 \pm 0.006$ \\
\nodata & \nodata& \nodata  &\nodata & $15,415\pm  34$\tablenotemark{b} 
                         & $3.365\pm 0.014$\tablenotemark{b}  
                   &  $114.0\pm 2.6$\tablenotemark{b} 
             & $0.798\pm 0.002$\tablenotemark{b}  \\
\hline
2007-12-21\tablenotemark{c} & 4455.56375 & 0.333 & $0.86\pm0.78$         & \nodata         & \nodata        & \nodata          & \nodata           \\
2007-12-21\tablenotemark{c} & 4455.58585 & 0.346 & $0.86\pm0.78$         & \nodata         & \nodata        & \nodata          & \nodata           \\
2007-12-21\tablenotemark{c} & 4455.63073 & 0.372 & $0.86\pm0.78$         & \nodata         & \nodata        & \nodata          & \nodata           \\
2007-12-21    &        4455.65994   & 0.389     &            $0.86\pm0.70$ &$ 15,872\pm   120$  &  $3.247 \pm 0.022 $  &  $114.4 \pm  4.9$ & $ 0.914 \pm 0.007$ \\
2007-12-21    &        4455.68207   & 0.402     &            $0.86\pm0.70$ &$ 16,048\pm   151$  &  $3.294 \pm 0.025 $  &  $120.8 \pm  5.1$ & $ 0.885 \pm 0.006$ \\
2007-12-21    &        4455.70422   & 0.415     &            $0.86\pm0.70$ &$ 16,164\pm   235$  &  $3.239 \pm 0.032 $  &  $111.4 \pm  7.6$ & $ 0.886 \pm 0.011$ \\
2007-12-21    &        4455.72745   & 0.429     &            $0.86\pm0.70$ &$ 15,817\pm   160$  &  $3.322 \pm 0.020 $  &  $107.1 \pm  4.1$ & $ 0.851 \pm 0.005$ \\
2007-12-21    &        4455.74956   & 0.442     &            $0.86\pm0.70$ &$ 16,114\pm   139$  &  $3.287 \pm 0.017 $  &  $106.9 \pm  5.9$ & $ 0.815 \pm 0.006$ \\
2007-12-21    &        4455.77165   & 0.455     &            $0.86\pm0.70$ &$ 16,498\pm   203$  &  $3.288 \pm 0.024 $  &  $116.8 \pm  4.9$ & $ 0.837 \pm 0.007$ \\
2007-12-21    &        4455.79313   & 0.467     &            $0.86\pm0.70$ &$ 16,984\pm   141$  &  $3.391 \pm 0.020 $  &  $117.2 \pm  3.3$ & $ 0.805 \pm 0.005$ \\
2007-12-21    &        4455.81523   & 0.480     &            $0.86\pm0.70$ &$ 16,825\pm   145$  &  $3.436 \pm 0.023 $  &  $130.1 \pm  5.2$ & $ 0.824 \pm 0.006$ \\
2007-12-21    &        4455.83735   & 0.493     &            $0.86\pm0.70$ &$ 15,958\pm   233$  &  $3.356 \pm 0.041 $  &  $114.5 \pm  4.9$ & $ 0.836 \pm 0.007$ \\
\nodata & \nodata& \nodata  &\nodata & $18,085\pm  93$\tablenotemark{b} 
                         & $3.297\pm 0.009$\tablenotemark{b}  
                   &  $120.2\pm 2.2$\tablenotemark{b} 
             & $0.800\pm 0.004$\tablenotemark{b}  \\
\hline
2008-02-27    &        4523.54114   & 0.207     &            $2.21\pm0.63$ &$ 15,498\pm    85$  &  $3.000 \pm 0.015 $  &  $167.7 \pm 11.2$ & $ 0.766 \pm 0.009$ \\
2008-02-27    &        4523.56408   & 0.220     &            $2.21\pm0.63$ &$ 15,399\pm   249$  &  $2.984 \pm 0.018 $  &  $184.5 \pm  6.7$ & $ 0.769 \pm 0.013$ \\
2008-02-27    &        4523.58682   & 0.233     &            $2.21\pm0.63$ &$ 15,399\pm    55$  &  $2.984 \pm 0.006 $  &  $184.5 \pm  2.5$ & $ 0.747 \pm 0.003$ \\
2008-02-27    &        4523.60952   & 0.247     &            $2.21\pm0.63$ &$ 15,717\pm   146$  &  $2.898 \pm 0.058 $  &  $193.0 \pm  2.7$ & $ 0.785 \pm 0.012$ \\
2008-02-27    &        4523.63193   & 0.260     &            $2.21\pm0.63$ &$ 15,704\pm   151$  &  $2.955 \pm 0.057 $  &  $188.2 \pm  6.8$ & $ 0.760 \pm 0.007$ \\
2008-02-27    &        4523.67602   & 0.286     &            $2.21\pm0.63$ &$ 15,653\pm   353$  &  $3.048 \pm 0.055 $  &  $162.9 \pm  8.7$ & $ 0.797 \pm 0.017$ \\
\nodata & \nodata& \nodata  &\nodata & $17,003\pm  21$\tablenotemark{b} 
                         & $3.415\pm 0.007$\tablenotemark{b}  
                   &  $160.3\pm 2.3$\tablenotemark{b} 
             & $0.717\pm 0.002$\tablenotemark{b}  \\
\hline
2008-02-28    &        4524.53912   & 0.792     &            $2.24\pm0.39$ &$ 16,302\pm   122$  &  $3.793 \pm 0.020 $  &  $111.3 \pm  5.3$ & $ 0.736 \pm 0.003$ \\
2008-02-28    &        4524.56150   & 0.805     &            $2.24\pm0.39$ &$ 15,677\pm   145$  &  $3.681 \pm 0.025 $  &  $121.4 \pm  6.6$ & $ 0.723 \pm 0.006$ \\
2008-02-28    &        4524.58384   & 0.818     &            $2.24\pm0.39$ &$ 15,357\pm   116$  &  $3.801 \pm 0.024 $  &  $100.9 \pm  5.2$ & $ 0.720 \pm 0.004$ \\
2008-02-28    &        4524.60626   & 0.831     &            $2.24\pm0.39$ &$ 15,895\pm    84$  &  $3.849 \pm 0.023 $  &  $ 94.9 \pm  4.6$ & $ 0.731 \pm 0.004$ \\
2008-02-28    &        4524.62862   & 0.845     &            $2.24\pm0.39$ &$ 15,808\pm    62$  &  $3.630 \pm 0.020 $  &  $103.9 \pm  3.5$ & $ 0.705 \pm 0.002$ \\
2008-02-28    &        4524.65381   & 0.859     &            $2.24\pm0.39$ &$ 15,727\pm   115$  &  $3.522 \pm 0.021 $  &  $108.0 \pm  3.7$ & $ 0.693 \pm 0.004$ \\
\nodata & \nodata& \nodata  &\nodata & $16,094\pm  71$\tablenotemark{b} 
                         & $3.435\pm 0.010$\tablenotemark{b}  
                   &  $106.3\pm 2.1$\tablenotemark{b} 
             & $0.661\pm 0.002$\tablenotemark{b}  \\
\hline
2008-02-29    &        4525.53557   & 0.377     &            $2.22\pm0.44$ &$ 16,536\pm   173$  &  $3.245 \pm 0.021 $  &  $118.0 \pm  3.4$ & $ 0.641 \pm 0.005$ \\
2008-02-29    &        4525.55787   & 0.390     &            $2.22\pm0.44$ &$ 16,650\pm   195$  &  $3.400 \pm 0.020 $  &  $108.9 \pm  6.8$ & $ 0.634 \pm 0.006$ \\
2008-02-29    &        4525.58156   & 0.406     &            $2.22\pm0.44$ &$ 16,595\pm   128$  &  $3.417 \pm 0.026 $  &  $106.2 \pm  5.3$ & $ 0.618 \pm 0.004$ \\
2008-02-29    &        4525.60393   & 0.417     &            $2.22\pm0.44$ &$ 16,850\pm   150$  &  $3.297 \pm 0.027 $  &  $118.0 \pm  4.3$ & $ 0.624 \pm 0.004$ \\
2008-02-29    &        4525.62682   & 0.431     &            $2.22\pm0.44$ &$ 16,650\pm    85$  &  $3.383 \pm 0.025 $  &  $114.9 \pm  3.3$ & $ 0.617 \pm 0.002$ \\
\nodata & \nodata& \nodata  &\nodata & $18,352\pm  73$\tablenotemark{b} 
                         & $3.522\pm 0.022$\tablenotemark{b}  
                   &  $107.6\pm 1.7$\tablenotemark{b} 
             & $0.565\pm 0.003$\tablenotemark{b}  \\
\hline
2008-03-01    &        4526.52329   & 0.956     &            $2.20\pm0.19$ &$ 15,996\pm    69$  &  $3.482 \pm 0.018 $  &  $122.4 \pm  3.5$ & $ 0.677 \pm 0.002$ \\
2008-03-01    &        4526.54550   & 0.969     &            $2.20\pm0.19$ &$ 15,440\pm    98$  &  $3.600 \pm 0.017 $  &  $120.3 \pm  3.9$ & $ 0.707 \pm 0.003$ \\
2008-03-01    &        4526.56776   & 0.982     &            $2.20\pm0.19$ &$ 15,653\pm   124$  &  $3.488 \pm 0.032 $  &  $147.0 \pm 11.9$ & $ 0.722 \pm 0.007$ \\
2008-03-01    &        4526.59005   & 0.995     &            $2.20\pm0.19$ &$ 15,545\pm   205$  &  $3.501 \pm 0.023 $  &  $126.6 \pm  5.5$ & $ 0.722 \pm 0.005$ \\
2008-03-01    &        4526.61241   & 0.008     &            $2.20\pm0.19$ &$ 15,475\pm   107$  &  $3.455 \pm 0.020 $  &  $140.3 \pm  5.3$ & $ 0.720 \pm 0.004$ \\
2008-03-01    &        4526.63475   & 0.021     &            $2.20\pm0.19$ &$ 15,461\pm    87$  &  $3.414 \pm 0.021 $  &  $125.3 \pm  3.9$ & $ 0.703 \pm 0.003$ \\
2008-03-01    &        4526.65820   & 0.035     &            $2.20\pm0.19$ &$ 15,168\pm    54$  &  $3.333 \pm 0.007 $  &  $139.3 \pm  7.4$ & $ 0.680 \pm 0.002$ \\
\nodata & \nodata& \nodata  &\nodata & $16,106\pm  105$\tablenotemark{b} 
                         & $3.241\pm 0.011$\tablenotemark{b}  
                   &  $125.0\pm 2.4$\tablenotemark{b} 
             & $0.634\pm 0.003$\tablenotemark{b}  \\
\enddata
\tablenotetext{a}{X-ray intensity from the {\em RXTE} 
All Sky Monitor.}
\tablenotetext{b}{Derived from averaged spectrum.}
\tablenotetext{c}{Contaminated by Balmer emission.}
\label{BAT}
\end{deluxetable}

\clearpage

\begin{deluxetable}{|l|rr|rr|rr|}
\setcounter{table}{4}
\rotate
\tablecaption{ELC model fits\label{tab3a}}
\tabletypesize{\footnotesize}
\tablewidth{0pt}
\tablehead{
\colhead{Parameter} &
\colhead{A93} &
\colhead{A93} &
\colhead{W94} &
\colhead{W94} &
\colhead{W95} &
\colhead{W95} \\
\colhead{} &
\colhead{No heating} &
\colhead{X-ray heating} &
\colhead{No heating} &
\colhead{X-ray heating} &
\colhead{No heating} &
\colhead{X-ray heating} }
\startdata
$i$ (deg)                & $71.58 \pm  0.12$     & $72.29 \pm  0.42$     & $64.62 \pm  5.49$     & $62.88 \pm  4.40$     & $69.18 \pm  0.65$     & $67.43 \pm  1.27$      \cr
$\llog$   & \nodata & $37.75 \pm 0.13$  & \nodata & $38.50 \pm 0.07$       & \nodata & $38.06 \pm 0.44$ \cr 
$K_2$ (km s$^{-1}$)                & $242.02 \pm 2.75$     & $241.00 \pm 1.20$     & $241.11 \pm 2.14$     & $240.95 \pm 3.00$     & $241.21 \pm 3.00$     & $241.06 \pm 2.02$      \cr
$Q$                 & $2.31 \pm 0.08$     & $1.93 \pm 0.15$     & $1.93 \pm 0.23$     & $1.93 \pm 0.24$     & $1.97 \pm 0.16$     & $1.92 \pm 0.16$      \cr
$\Delta\phi$                 & $0.00084 \pm 0.00152$     & $0.00347 \pm 0.00248$     & $-0.04000 \pm 0.00828$     & $-0.02906 \pm 0.00800$     & $0.00046 \pm 0.00273$     & $-0.00309 \pm 0.00679$      \cr
$T_{\rm disk} (K)$                 & $42160.0 \pm    13.8$     & $36353.5 \pm 10826.2$     & $18423.7 \pm 15045.5$     & $22754.2 \pm  7144.2$     & $24395.1 \pm 11113.9$     & $17886.4 \pm  2145.7$      \cr
$r_{\rm out} $                 & $0.680 \pm 0.054$     & $0.400 \pm 0.053$     & $0.556 \pm 0.269$     & $0.694 \pm 0.175$     & $0.459 \pm 0.064$     & $0.503 \pm 0.118$      \cr
$\xi$                 & $-0.0062 \pm 0.0001$     & $-0.1118 \pm 0.0478$     & $-0.0959 \pm 0.0731$     & $-0.0698 \pm 0.0643$     & $-0.1200 \pm 0.0653$     & $-0.0601 \pm 0.0352$      \cr
$\beta$ (deg)                 & $ 5.00 \pm  0.51$     & $ 4.53 \pm  1.48$     & $ 4.46 \pm  3.29$     & $ 2.84 \pm  0.80$     & $ 1.50 \pm  2.01$     & $ 3.37 \pm  2.00$      \cr
$s_{\rm spot}$                 & $ 1.10 \pm  0.22$     & $ 3.59 \pm  1.87$     & $ 5.25 \pm  4.55$     & $ 5.13 \pm  4.54$     & $ 2.24 \pm  0.90$     & $ 3.88 \pm  5.34$      \cr
$\theta_{\rm spot}$ (deg)         & $ 30.02 \pm   3.42$     & $ 32.28 \pm   6.61$     & $208.32 \pm   9.31$     & $154.04 \pm   4.28$     & $185.84 \pm   3.24$     & $274.59 \pm 122.15$      \cr
$r_{\rm cut}$         & $0.562 \pm 0.067$     & $0.844 \pm 0.344$     & $0.845 \pm 0.344$     & $0.617 \pm 0.357$     & $0.702 \pm 0.267$     & $0.874 \pm 0.365$      \cr
$w_{\rm spot}$ (deg)        & $12.43 \pm  1.52$     & $ 8.11 \pm  4.96$     & $ 3.29 \pm 10.11$     & $ 5.32 \pm  4.59$     & $49.75 \pm  4.65$     & $24.43 \pm 24.85$      \cr
$B$ disk fraction& $0.40 \pm 0.02$  & $0.20 \pm 0.04$  & $0.32 \pm 0.17$  
& $0.37 \pm 0.10$  & $0.17 \pm 0.10$  & $0.25 \pm 0.08$  \cr
$V$ disk fraction& $0.51 \pm 0.02$  & $0.28 \pm 0.05$  & $0.18 \pm 0.09$  & $0.27 \pm 0.10$  & $0.15 \pm 0.03$  & $0.18 \pm 0.03$  \cr
$M_2$  $(M_{\sun})$& $2.60 \pm 0.15$& $3.41 \pm 0.59$& $4.00 \pm 1.03$& $4.18 \pm 0.89$& $3.51 \pm 0.45$& $3.80 \pm 0.61$ \cr
$R_2$ $(R_{\sun})$& $3.80 \pm 0.08$& $4.18 \pm 0.24$& $4.41 \pm 0.36$& $4.48 \pm 0.31$& $4.22 \pm 0.19$& $4.34 \pm 0.24$ \cr
$\log g_2$ (cgs)& $3.69 \pm 0.01$& $3.73 \pm 0.02$& $3.75 \pm 0.03$& $3.76 \pm 0.03$& $3.73 \pm 0.02$& $3.74 \pm 0.02$  \cr
$M$ ($M_{\sun}$)& $ 6.02 \pm 0.21$& $ 6.59 \pm 0.29$& $ 7.73 \pm 1.55$& $ 8.07 \pm 1.52$& $ 6.91 \pm 0.38$& $ 7.28 \pm 0.47$ \cr
\enddata
\end{deluxetable}
\clearpage

\begin{deluxetable}{|l|rr|rr|rr|}
\setcounter{table}{5}
\rotate
\tablecaption{ELC model fits\label{tab3b}}
\tabletypesize{\footnotesize}
\tablewidth{0pt}
\tablehead{
\colhead{Parameter} &
\colhead{S96} &
\colhead{S96} &
\colhead{A96a} &
\colhead{A96a} &
\colhead{A96b} &
\colhead{A96b} \\
\colhead{} &
\colhead{No heating} &
\colhead{X-ray heating} &
\colhead{No heating} &
\colhead{X-ray heating} &
\colhead{No heating} &
\colhead{X-ray heating} }
\startdata
$i$ (deg)                & $71.60 \pm  2.47$     & $70.46 \pm  0.93$     & $70.13 \pm  1.04$     & $67.84 \pm  0.66$     & $70.95 \pm  2.51$     & $70.63 \pm  0.71$      \cr
$\llog$   & \nodata & $38.21 \pm 0.11$  & \nodata & $36.00 \pm 1.18$       & \nodata & $36.03 \pm 1.00$ \cr 
$K_2$ (km s$^{-1}$)                & $241.02 \pm 3.01$     & $241.01 \pm 2.84$     & $241.25 \pm 1.68$     & $241.22 \pm 3.00$     & $241.66 \pm 3.44$     & $241.10 \pm 3.02$      \cr
$Q$                 & $1.90 \pm 0.18$     & $1.90 \pm 0.16$     & $2.06 \pm 0.20$     & $1.97 \pm 0.11$     & $1.98 \pm 0.16$     & $1.93 \pm 0.17$      \cr
$\Delta\phi$                 & $-0.00890 \pm 0.00575$     & $-0.01511 \pm 0.00480$     & $-0.01010 \pm 0.00534$     & $-0.01800 \pm 0.00307$     & $-0.02351 \pm 0.00391$     & $0.00230 \pm 0.00407$      \cr
$T_{\rm disk} (K)$                 & $46687.8 \pm 23161.4$     & $49247.9 \pm  2450.0$     & $35888.8 \pm 13316.2$     & $41286.1 \pm  7014.0$     & $33577.9 \pm  6930.0$     & $37509.2 \pm  3468.3$      \cr
$r_{\rm out} $                 & $0.607 \pm 0.066$     & $0.730 \pm 0.036$     & $0.408 \pm 0.217$     & $0.740 \pm 0.040$     & $0.737 \pm 0.161$     & $0.633 \pm 0.066$      \cr
$\xi$                 & $-0.1946 \pm 0.0888$     & $-0.2485 \pm 0.0334$     & $-0.1416 \pm 0.0824$     & $-0.1751 \pm 0.0241$     & $-0.2116 \pm 0.0169$     & $-0.1737 \pm 0.0252$      \cr
$\beta$ (deg)                 & $ 1.86 \pm  2.71$     & $ 1.62 \pm  0.72$     & $ 1.01 \pm  0.15$     & $ 2.57 \pm  1.49$     & $ 1.32 \pm  0.78$     & $ 1.51 \pm  0.75$      \cr
$s_{\rm spot}$                 & $ 1.66 \pm  1.71$     & $ 3.88 \pm  1.82$     & $ 2.00 \pm  1.14$     & $ 4.53 \pm  4.45$     & $ 3.34 \pm  6.05$     & $ 4.80 \pm  4.45$      \cr
$\theta_{\rm spot}$ (deg)         & $190.14 \pm  11.33$     & $351.76 \pm   7.95$     & $185.79 \pm   6.88$     & $  4.19 \pm   3.20$     & $144.24 \pm   5.56$     & $338.06 \pm   7.79$      \cr
$r_{\rm cut}$         & $0.500 \pm 0.488$     & $0.618 \pm 0.371$     & $0.793 \pm 0.228$     & $0.724 \pm 0.260$     & $0.764 \pm 0.261$     & $0.615 \pm 0.368$      \cr
$w_{\rm spot}$ (deg)        & $47.43 \pm 37.68$     & $13.74 \pm 19.70$     & $49.91 \pm  9.78$     & $ 8.83 \pm  2.80$     & $21.61 \pm 26.63$     & $11.46 \pm 23.27$      \cr
$B$ disk fraction& $0.24 \pm 0.07$  & $0.30 \pm 0.03$  & $0.13 \pm 0.11$  & $0.36 \pm 0.04$  & $0.30 \pm 0.09$  & $0.25 \pm 0.04$  \cr
$V$ disk fraction& $0.25 \pm 0.08$  & $0.15 \pm 0.02$  & $0.15 \pm 0.01$  & $0.29 \pm 0.03$  & $0.15 \pm 0.02$  & $0.16 \pm 0.02$  \cr
$M_2$  $(M_{\sun})$& $3.54 \pm 0.49$& $3.62 \pm 0.52$& $3.20 \pm 0.59$& $3.61 \pm 0.40$& $3.36 \pm 0.57$& $3.53 \pm 0.65$ \cr
$R_2$ $(R_{\sun})$& $4.24 \pm 0.22$& $4.27 \pm 0.23$& $4.09 \pm 0.25$& $4.26 \pm 0.16$& $4.16 \pm 0.23$& $4.23 \pm 0.29$ \cr
$\log g_2$ (cgs)& $3.73 \pm 0.02$& $3.74 \pm 0.02$& $3.72 \pm 0.02$& $3.74 \pm 0.01$& $3.73 \pm 0.02$& $3.73 \pm 0.03$  \cr
$M$ ($M_{\sun}$)& $ 6.74 \pm 0.51$& $ 6.88 \pm 0.44$& $ 6.59 \pm 0.47$& $ 7.10 \pm 0.42$& $ 6.68 \pm 0.59$& $ 6.80 \pm 0.40$ \cr
\enddata
\end{deluxetable}

\clearpage

\begin{deluxetable}{|l|rr|rr|rr|}
\setcounter{table}{6}
\rotate
\tablecaption{ELC model fits\label{tab3c}}
\tabletypesize{\footnotesize}
\tablewidth{0pt}
\tablehead{
\colhead{Parameter} &
\colhead{W96a} &
\colhead{W96a} &
\colhead{W96b} &
\colhead{W96b} &
\colhead{S98} &
\colhead{S98} \\
\colhead{} &
\colhead{No heating} &
\colhead{X-ray heating} &
\colhead{No heating} &
\colhead{X-ray heating} &
\colhead{No heating} &
\colhead{X-ray heating}}
\startdata
$i$ (deg)                & $69.24 \pm  1.83$     & $69.17 \pm  1.19$     & $65.73 \pm  1.22$     & $63.71 \pm  3.29$     & $60.61 \pm  6.21$     & $63.61 \pm  0.75$      \cr
$\llog$   & \nodata & $37.64 \pm 1.50$  & \nodata & $38.31 \pm 1.62$       & \nodata & $38.05 \pm 0.09$ \cr 
$K_2$ (km s$^{-1}$)                & $241.10 \pm 2.92$     & $241.08 \pm 3.00$     & $241.51 \pm 3.17$     & $240.71 \pm 4.29$     & $241.40 \pm 1.91$     & $241.01 \pm 1.94$      \cr
$Q$                 & $1.93 \pm 0.14$     & $1.92 \pm 0.16$     & $1.83 \pm 0.15$     & $1.85 \pm 0.43$     & $2.02 \pm 0.24$     & $1.90 \pm 0.13$      \cr
$\Delta\phi$                 & $-0.01546 \pm 0.00502$     & $-0.01626 \pm 0.00640$     & $0.02484 \pm 0.01040$     & $0.01864 \pm 0.01523$     & $-0.01059 \pm 0.00676$     & $-0.01363 \pm 0.00391$      \cr
$T_{\rm disk} (K)$                 & $39974.9 \pm 10500.0$     & $43156.5 \pm  3500.1$     & $46351.3 \pm   913.3$     & $28674.6 \pm 20437.1$     & $37757.6 \pm  8943.4$     & $41497.6 \pm 26095.5$      \cr
$r_{\rm out} $                 & $0.966 \pm 0.073$     & $0.921 \pm 0.053$     & $0.969 \pm 0.052$     & $0.932 \pm 0.057$     & $0.875 \pm 0.387$     & $0.746 \pm 0.043$      \cr
$\xi$                 & $-0.2959 \pm 0.0198$     & $-0.3086 \pm 0.0228$     & $-0.0488 \pm 0.0029$     & $-0.0710 \pm 0.0933$     & $-0.1789 \pm 0.1631$     & $-0.2121 \pm 0.1690$      \cr
$\beta$ (deg)                 & $ 4.83 \pm  3.01$     & $ 5.00 \pm  1.20$     & $ 1.17 \pm  1.10$     & $ 4.55 \pm  2.88$     & $ 4.75 \pm  3.24$     & $ 4.99 \pm  1.06$      \cr
$s_{\rm spot}$                 & $ 5.91 \pm  0.67$     & $ 5.68 \pm  0.86$     & $ 3.34 \pm  2.12$     & $ 1.69 \pm  5.21$     & $ 0.80 \pm  0.88$     & $ 1.40 \pm  0.22$      \cr
$\theta_{\rm spot}$ (deg)         & $191.15 \pm   5.13$     & $192.83 \pm   6.51$     & $332.44 \pm  16.14$     & $ 35.01 \pm 322.31$     & $ 21.11 \pm   8.60$     & $322.48 \pm  10.27$      \cr
$r_{\rm cut}$         & $0.500 \pm 0.197$     & $0.506 \pm 0.130$     & $0.517 \pm 0.206$     & $0.869 \pm 0.364$     & $0.940 \pm 0.433$     & $0.629 \pm 0.339$      \cr
$w_{\rm spot}$ (deg)        & $25.76 \pm 16.75$     & $30.35 \pm  7.08$     & $25.97 \pm 23.65$     & $ 3.80 \pm 46.18$     & $48.70 \pm  7.55$     & $15.12 \pm 13.87$      \cr
$B$ disk fraction& $0.50 \pm 0.07$  & $0.49 \pm 0.07$  & $0.44 \pm 0.02$  & $0.51 \pm 0.08$  & $0.52 \pm 0.29$  & $0.43 \pm 0.03$  \cr
$V$ disk fraction& $0.18 \pm 0.05$  & $0.19 \pm 0.03$  & $0.69 \pm 0.14$  & $0.56 \pm 0.12$  & $0.38 \pm 0.08$  & $0.28 \pm 0.04$  \cr
$M_2$  $(M_{\sun})$& $3.62 \pm 0.43$& $3.64 \pm 0.43$& $4.29 \pm 0.59$& $4.39 \pm 1.10$& $4.15 \pm 0.35$& $4.21 \pm 0.45$ \cr
$R_2$ $(R_{\sun})$& $4.27 \pm 0.17$& $4.28 \pm 0.17$& $4.53 \pm 0.23$& $4.56 \pm 0.43$& $4.46 \pm 0.14$& $4.49 \pm 0.17$ \cr
$\log g_2$ (cgs)& $3.74 \pm 0.01$& $3.74 \pm 0.02$& $3.76 \pm 0.02$& $3.76 \pm 0.04$& $3.76 \pm 0.01$& $3.76 \pm 0.01$  \cr
$M$ ($M_{\sun}$)& $ 6.99 \pm 0.43$& $ 7.01 \pm 0.39$& $ 7.86 \pm 0.54$& $ 8.11 \pm 1.30$& $ 8.39 \pm 0.99$& $ 8.01 \pm 0.36$ \cr
\enddata
\end{deluxetable}

\clearpage

\begin{deluxetable}{|l|rr|rr|rr|}
\setcounter{table}{7}
\rotate
\tablecaption{ELC model fits\label{tab3d}}
\tabletypesize{\footnotesize}
\tablewidth{0pt}
\tablehead{
\colhead{Parameter} &
\colhead{W98} &
\colhead{W98} &
\colhead{vdK} &
\colhead{vdK} &
\colhead{Halved} &
\colhead{Halved} \\
\colhead{} &
\colhead{No heating} &
\colhead{X-ray heating} &
\colhead{No heating} &
\colhead{X-ray heating} &
\colhead{No heating} &
\colhead{X-ray heating}}
\startdata
$i$ (deg)                & $66.17 \pm  1.86$     & $63.34 \pm  1.40$     & $69.54 \pm  0.46$     & $69.80 \pm  1.29$     & $70.02 \pm  0.40$     & $69.79 \pm  0.22$      \cr
$\llog$  & \nodata & $38.13 \pm 0.34$  & \nodata & $37.85 \pm 0.26$       & \nodata & $37.60 \pm 0.23$ \cr 
$K_2$ (km s$^{-1}$)                & $240.13 \pm 2.85$     & $241.04 \pm 3.00$     & $241.11 \pm 2.96$     & $241.03 \pm 2.99$     & $242.55 \pm 1.93$     & $239.55 \pm 3.62$      \cr
$Q$                 & $1.98 \pm 0.19$     & $1.91 \pm 0.24$     & $1.94 \pm 0.16$     & $1.91 \pm 0.16$     & $2.24 \pm 0.14$     & $2.13 \pm 0.10$      \cr
$\Delta\phi$                 & $-0.00145 \pm 0.00285$     & $-0.00195 \pm 0.00523$     & $0.02702 \pm 0.00160$     & $0.02704 \pm 0.00232$     & $-0.00073 \pm 0.00070$     & $-0.00294 \pm 0.00402$      \cr
$T_{\rm disk} (K)$                 & $20803.0 \pm  5638.1$     & $22031.7 \pm  2540.2$     & $30036.7 \pm  7015.6$     & $42288.4 \pm  7365.4$     & $30084.9 \pm   172.8$     & $27732.5 \pm 12523.5$      \cr
$r_{\rm out} $                 & $0.990 \pm 0.026$     & $0.990 \pm 0.021$     & $0.901 \pm 0.039$     & $0.912 \pm 0.059$     & $0.401 \pm 0.000$     & $0.405 \pm 0.008$      \cr
$\xi$                 & $-0.1474 \pm 0.0293$     & $-0.1569 \pm 0.0101$     & $-0.1613 \pm 0.0261$     & $-0.1867 \pm 0.1031$     & $-0.1638 \pm 0.0278$     & $-0.1133 \pm 0.0811$      \cr
$\beta$ (deg)                 & $ 2.37 \pm  2.16$     & $ 3.69 \pm  1.20$     & $ 4.24 \pm  1.02$     & $ 2.14 \pm  2.67$     & $ 4.46 \pm  2.32$     & $ 2.25 \pm  0.88$      \cr
$s_{\rm spot}$                 & $ 1.81 \pm  3.33$     & $ 6.09 \pm  1.31$     & $ 1.27 \pm  0.08$     & $ 2.59 \pm  1.07$     & $ 1.36 \pm  0.25$     & $ 1.57 \pm  0.57$      \cr
$\theta_{\rm spot}$ (deg)         & $191.88 \pm   4.45$     & $128.52 \pm  46.82$     & $178.71 \pm   3.80$     & $312.57 \pm 307.65$     & $173.81 \pm   2.46$     & $130.44 \pm  47.40$      \cr
$r_{\rm cut}$         & $0.754 \pm 0.236$     & $0.749 \pm 0.248$     & $0.983 \pm 0.260$     & $0.639 \pm 0.293$     & $0.787 \pm 0.203$     & $0.708 \pm 0.201$      \cr
$w_{\rm spot}$ (deg)        & $49.21 \pm 39.31$     & $ 1.44 \pm 45.45$     & $49.14 \pm  5.01$     & $ 5.89 \pm 43.42$     & $50.00 \pm  1.89$     & $14.68 \pm 19.01$      \cr
$B$ disk fraction& $0.49 \pm 0.07$  & $0.52 \pm 0.03$  & $0.46 \pm 0.03$  & $0.41 \pm 0.08$  & $0.23 \pm 0.06$  & $0.17 \pm 0.03$  \cr
$V$ disk fraction& $0.19 \pm 0.08$  & $0.15 \pm 0.03$  & $0.28 \pm 0.03$  & $0.31 \pm 0.03$  & $0.16 \pm 0.01$  & $0.15 \pm 0.00$  \cr
$M_2$  $(M_{\sun})$& $3.66 \pm 0.72$& $4.22 \pm 0.75$& $3.58 \pm 0.43$& $3.64 \pm 0.49$& $2.84 \pm 0.35$& $2.97 \pm 0.33$ \cr
$R_2$ $(R_{\sun})$& $4.28 \pm 0.28$& $4.49 \pm 0.30$& $4.25 \pm 0.19$& $4.28 \pm 0.21$& $3.92 \pm 0.18$& $3.98 \pm 0.15$ \cr
$\log g_2$ (cgs)& $3.74 \pm 0.02$& $3.76 \pm 0.03$& $3.73 \pm 0.02$& $3.74 \pm 0.02$& $3.70 \pm 0.02$& $3.71 \pm 0.01$  \cr
$M$ ($M_{\sun}$)& $ 7.24 \pm 0.76$& $ 8.05 \pm 0.61$& $ 6.93 \pm 0.41$& $ 6.95 \pm 0.44$& $ 6.36 \pm 0.09$& $ 6.34 \pm 0.33$ \cr
\enddata
\end{deluxetable}

\clearpage

\begin{deluxetable}{|l|rr|}
\setcounter{table}{8}
\tablecaption{ELC model fits\label{tab3e}}
\tabletypesize{\footnotesize}
\tablewidth{0pt}
\tablehead{
\colhead{Parameter} &
\colhead{Quartered} &
\colhead{Quartered} \\
\colhead{} &
\colhead{No heating} &
\colhead{X-ray heating} }
\startdata
$i$ (deg)                & $69.52 \pm  0.24$      
& $68.92 \pm  0.61$      \cr
$\llog$   & \nodata & $37.59 \pm 0.87$ \cr 
$K_2$ (km s$^{-1}$)                & $241.20 \pm 3.02$      
& $241.08 \pm 3.00$      \cr
$Q$                 & $1.96 \pm 0.04$      
& $1.93 \pm 0.20$      \cr
$\Delta\phi$                 & $0.00614 \pm 0.01014$      
& $0.01404 \pm 0.01383$      \cr
$T_{\rm disk} (K)$                 & $44931.6 \pm    17.5$ 
& $36366.6 \pm  1520.0$      \cr
$r_{\rm out} $                 & $0.400 \pm 0.001$     
& $0.443 \pm 0.130$      \cr
$\xi$                 & $-0.0785 \pm 0.0037$      
& $-0.0163 \pm 0.0030$      \cr
$\beta$ (deg)                 & $ 1.98 \pm  0.47$   
& $ 3.65 \pm  2.34$      \cr
$s_{\rm spot}$                 & $ 7.94 \pm  0.89$  
& $ 1.84 \pm  0.78$      \cr
$\theta_{\rm spot}$ (deg)         & $  0.36 \pm   6.84$ 
& $326.49 \pm 242.46$      \cr
$r_{\rm cut}$         & $0.575 \pm 0.279$  
& $0.552 \pm 0.435$      \cr
$w_{\rm spot}$ (deg)        & $44.80 \pm 12.10$ 
& $ 8.96 \pm 21.12$      \cr
$B$ disk fraction& $0.15 \pm 0.01$ 
& $0.22 \pm 0.04$  \cr
$V$ disk fraction& $0.32 \pm 0.02$ 
& $0.21 \pm 0.04$  \cr
$M_2$  $(M_{\sun})$& $3.51 \pm 0.13$ 
& $3.65 \pm 0.47$ \cr
$R_2$ $(R_{\sun})$& $4.22 \pm 0.05$ 
& $4.28 \pm 0.21$ \cr
$\log g_2$ (cgs)& $3.73 \pm 0.01$  
& $3.74 \pm 0.02$  \cr
$M$ ($M_{\sun}$)& $ 6.88 \pm 0.25$ 
& $ 7.03 \pm 0.44$ \cr
\enddata
\end{deluxetable}

\end{document}